\documentclass[preprint,12pt]{aastex}
\usepackage{natbib}
\usepackage{graphicx}

\begin{document}

\title{The Mid-Infrared Instrument for JWST, II: Design and Build}

\author
{G. S. Wright\altaffilmark{1},
David Wright\altaffilmark{2},
G. B. Goodson\altaffilmark{3}, 
G. H. Rieke\altaffilmark{4}, 
Gabby Aitink-Kroes\altaffilmark{5},
J. Amiaux\altaffilmark{6},
Ana Aricha-Yanguas\altaffilmark{7},
Ruym\'an Azzolini\altaffilmark{8,9},
Kimberly Banks\altaffilmark{10},
D.  Barrado-Navascues\altaffilmark{9},
T. Belenguer-Davila\altaffilmark{7},
J. A. D. L. Bloemmart\altaffilmark{11,12,13},
Patrice Bouchet\altaffilmark{6},
B. R. Brandl\altaffilmark{14},
L. Colina\altaffilmark{9},
\"{O}rs Detre\altaffilmark{15},
Eva Diaz-Catala\altaffilmark{7},
Paul Eccleston\altaffilmark{16},
Scott D. Friedman\altaffilmark{17},
Macarena Garc\'ia-Mar\'in\altaffilmark{18},
Manuel G\"udel\altaffilmark{19,20},
Alistair Glasse\altaffilmark{1},
Adrian M. Glauser\altaffilmark{20},
T. P. Greene\altaffilmark{21},
Uli Groezinger\altaffilmark{15},
Tim Grundy\altaffilmark{16},
Peter Hastings\altaffilmark{1},
Th. Henning\altaffilmark{15},
Ralph Hofferbert\altaffilmark{15},
Faye Hunter\altaffilmark{22},
N. C. Jessen\altaffilmark{23},
K. Justtanont\altaffilmark{24},
Avinash R. Karnik\altaffilmark{25},
Mori A. Khorrami\altaffilmark{3},
Oliver Krause\altaffilmark{15},
Alvaro Labiano\altaffilmark{20},
P.-O. Lagage\altaffilmark{6},
Ulrich Langer\altaffilmark{26},
Dietrich Lemke\altaffilmark{15},
Tanya Lim\altaffilmark{16},
Jose Lorenzo-Alvarez\altaffilmark{27},
Emmanuel Mazy\altaffilmark{28},
Norman McGowan\altaffilmark{22}, 
M. E. Meixner\altaffilmark{17, 29},
Nigel Morris\altaffilmark{16},
Jane E. Morrison\altaffilmark{4},
Friedrich M\"uller\altaffilmark{15},
H.-U. N\o rgaard-Nielson\altaffilmark{23},
G\"oran Olofsson\altaffilmark{24},
Brian O'Sullivan\altaffilmark{30},
J.-W. Pel\altaffilmark{31},
Konstantin Penanen\altaffilmark{3}, 
M. B. Petach\altaffilmark{32}, 
J. P. Pye\altaffilmark{33},
T. P. Ray\altaffilmark{8},
Etienne Renotte\altaffilmark{28},
Ian Renouf\altaffilmark{22},
M. E. Ressler\altaffilmark{3},
Piyal Samara-Ratna\altaffilmark{33},
Silvia Scheithauer\altaffilmark{15}, 
Analyn Schneider\altaffilmark{3},
Bryan Shaughnessy\altaffilmark{16},
Tim Stevenson\altaffilmark{34},
Kalyani Sukhatme\altaffilmark{3},
Bruce Swinyard\altaffilmark{16, 35},
Jon Sykes\altaffilmark{33},
John Thatcher\altaffilmark{36},
Tuomo Tikkanen\altaffilmark{33},
E. F. van Dishoeck\altaffilmark{14},
C. Waelkens\altaffilmark{11},
Helen Walker\altaffilmark{16},
Martyn Wells\altaffilmark{1},
Alex Zhender\altaffilmark{37}
}

\nopagebreak

\altaffiltext{1}{UK Astronomy Technology Centre, Royal Observatory,
  Blackford Hill Edinburgh, EH9 3HJ, Scotland, United Kingdom}
\altaffiltext{2}{Stinger Ghaffarian Technologies, Inc., Greenbelt, MD, USA.}
\altaffiltext{3}{Jet Propulsion Laboratory, California Institute of Technology, 4800 Oak Grove Dr. Pasadena, CA 91109, USA}
\altaffiltext{4}{Steward Observatory, 933 N. Cherry Ave, University of Arizona, Tucson, AZ 85721, USA}
\altaffiltext{5}{NOVA Opt-IR group, PO Box 2, 7990 AA Dwingeloo, The Netherlands}
\altaffiltext{6}{Laboratoire AIM Paris-Saclay, CEA-IRFU/SAp, CNRS, Université Paris Diderot, F-91191 Gif-sur-Yvette, France}
\altaffiltext{7}{INTA, Carretera de Ajalvir, km 4, 28850 Torrejon de Ardoz, Madrid, Spain}
\altaffiltext{8}{Dublin Institute for Advanced Studies, School of Cosmic Physics, 31 Fitzwilliam Place, Dublin 2, Ireland}
\altaffiltext{9}{Centro de Astrobiolog\'ia (INTA-CSIC), Dpto Astrof\'isica, Carretera de Ajalvir, km 4, 28850 Torrej\'on de Ardoz, Madrid, Spain}
\altaffiltext{10}{NASA Goddard Space Flight Ctr. , 8800 Greenbelt Rd., Greenbelt, MD 20771, USA}
\altaffiltext{11}{Institute of Astronomy KU Leuven, Celestijnenlaan 200D,3001 Leuven, Belgium}
\altaffiltext{12}{Astronomy and Astrophysics Research Group, Department of Physics and Astrophysics, Vrije Universiteit Brussel, Belgium}
\altaffiltext{13}{Flemish Institute for Technological Research (VITO), Boeretang 200,2400 Mol, Belgium}
\altaffiltext{14}{Leiden Observatory, Leiden University, PO Box 9513, 2300 RA, Leiden, The Netherlands.}
\altaffiltext{15}{Max Planck Institute f\"ur Astronomy (MPIA), K\"onigstuhl 17, D-69117 Heidelberg, Germany}
\altaffiltext{16}{RAL Space, STFC, Rutherford Appleton Lab., Harwell, Oxford, Didcot OX11 0QX, UK}
\altaffiltext{17}{Space Telescope Science Institute, 3700 San Martin Drive, Baltimore, MD 21218, USA}
\altaffiltext{18}{I. Physikalisches Institut, Universit\"at zu K\"oln, Z\"ulpicher Str. 77,  50937 K\"oln, Germany }
\altaffiltext{19}{Dept. of Astrophysics, Univ. of Vienna, T\"urkenschanzstr 17, A-1180 Vienna, Austria}
\altaffiltext{20}{ETH Zurich, Institute for Astronomy, Wolfgang-Pauli-Str. 27, CH-8093 Zurich, Switzerland}
\altaffiltext{21}{NASA Ames Research Center, M.S. 245-6, Moffett Field, CA 94035, USA}
\altaffiltext{22}{Airbus Defence and Space, Anchorage Road, Portsmouth, Hampshire, PO3 5PU}
\altaffiltext{23}{National Space Institute (DTU Space), Technical University of Denmark, Juliane Mariesvej 30, DK-2100, Copenhagen, Denmark}
\altaffiltext{24}{Chalmers University of Technology, Onsala Space Observatory, S-439 92 Onsala, Sweden}
\altaffiltext{3,25}{952 Camino Del Arroyo Dr., San Marcos, CA 92078, USA}
\altaffiltext{26}{RUAG Space, Schaffhauserstrasse 580, CH-8052 Z\"urich, Switzerland}
\altaffiltext{27}{ESTEC, Keplerlaan 1, 2201 AZ Noordwijk, The Netherlands}
\altaffiltext{28}{Centre Spatial De Li\'ege, Avenue du Pre Aily, B-4031, Angleur, Belguim}
\altaffiltext{29}{The Johns Hopkins University, Department of Physics and Astronomy, 366 Bloomberg Center, 3400 N. Charles Street, Baltimore, MD 21218, USA}
\altaffiltext{30}{Airbus Defence and Space, Anchorage Road, Portsmouth, Hampshire, PO3 5PU}
\altaffiltext{31}{Kapteyn Institute, University of Groningen, PO Box 800, 9700 Groningen, The Netherlands}
\altaffiltext{32}{Northrop-Grumman Aerospace Systems, One Space Park, Redondo Beach, CA 90278, USA}
\altaffiltext{33}{Department of Physics and Astronomy, Univ. of Leicester, University Road, Leicester, LE1 7RH, UK}
\altaffiltext{34}{SKA Organisation, Jodrell Bank Observatory, Lower Withington, Macclesfield, Cheshire, SK11 9DL, UK}
\altaffiltext{35}{Dept. Physics and Astronomy, University College London, Gower Place, London WC1E 6BT, London, UK}
\altaffiltext{36}{Airbus Defence and Space, Gunnels Wood Road, Stevenage, Hertfordshire, SG1, 2AS, UK}
\altaffiltext{37}{Paul Scherrer Institut, CH-5232 Villigen PSI, Switzerland }

\newpage

\begin{abstract}

The Mid-InfraRed Instrument (MIRI) on the James Webb 
Space Telescope (JWST) provides measurements over the wavelength range 5 to 
28.5 $\mu $m. MIRI has, within a single `package', four key scientific 
functions: photometric imaging, coronagraphy, single-source low-spectral 
resolving power (R $\sim$ 100) spectroscopy, and medium-resolving 
power (R $\sim$ 1500 to 3500) integral field spectroscopy. An 
associated cooler system maintains MIRI at its operating temperature of 
\textless 6.7 K. This paper describes the driving principles behind the 
design of MIRI, the primary design parameters, and their realisation in 
terms of the `as-built' instrument. It also describes the test programme 
that led to delivery of the tested and calibrated Flight Model to NASA in 
2012, and the confirmation after delivery of the key interface requirements.

\end{abstract}


\section{Introduction}
MIRI was the first of the JWST's main science instruments to be delivered to 
the NASA Goddard Spaceflight Center in the spring of 2012. That delivery 
marked a major milestone in the work of the consortium of European and US 
institutes (Rieke et al., 2015a, hereafter paper I) that had designed and 
built MIRI over a period of more than 10 years. This paper describes the 
overall instrument design and the development approach. It thereby provides 
the potential user of MIRI and its data with an insight into the engineering 
solutions that shape its operation and performance.

The MIRI instrument is the only mid-infrared instrument for JWST. To support 
a full range of investigations, it therefore provides four key scientific 
functions, whose detailed implementation is described elsewhere: 1) 
photometric imaging in nine wave-bands between 5 $\mu $m and 27 $\mu $m over 
a 2.3 square arcminute field of view (Bouchet et al., 2015, Paper III); 2) 
low spectral resolving power (R $\sim$ 100) spectroscopy of compact 
sources between 7 and 12 $\mu $m (Kendrew et al., 2015, Paper IV); 3) 
coronagraphy in 4 wave-bands between 10 $\mu $m and 27 $\mu $m (Boccaletti 
et al., 2015, Paper V); and 4) medium spectral resolution (R $\sim$ 
1500 to 3500) integral field spectroscopy over a 13 square arcsecond field 
of view between 5 and 28.5 $\mu $m (Wells et al., 2015, Paper VI). Each of 
these capabilities, coupled with the large, cold, aperture of JWST will 
provide a significant advance. To design all of them into a single 
instrument required novel designs and pushed manufacturing tolerances to the 
limits. In this paper we present the common design features of MIRI that 
support and enable these functions, and discuss how they were integrated 
into the delivered Flight Model. 

A total of three models of the MIRI instrument hardware were built, 
including the Flight Model (FM) shown in Figure 1 
and the subject of this paper and the accompanying ones. The Verification 
Model (VM) was fully operational (though with reduced imager and 
spectrometer functionality), and was built to de-risk the opto-mechanical 
concepts and assembly integration and verification programme. The first 
model, the Structural and Thermal Model (the STM), was built to be thermally 
and mechanically representative of the FM, to enable early validation of the 
thermal design and structural integrity. The STM has subsequently been 
enhanced with a representative focal plane so that it can be used in the 
development of the MIRI cooler. 

\section{Instrument architecture}
MIRI comprises two main components with associated assemblies: the MIRI 
Optical Bench Assembly (OBA) (Section 2.1) and the MIRI cooler system (Section 
5.2), which are operated via separate modules of the MIRI Flight Software 
running on the JWST Science instrument command and data handling system 
(ICDH).

\subsection{Optical Bench Assembly}

The OBA consists of the optics module (the OM, shown in 
Figure 1), the electrical control and data handling 
boxes associated with MIRI, which are maintained at 300 K in the separate 
ISIM Electronics Compartment (IEC; ISIM $=$ Integrated Science Instrument 
Module), and the necessary interconnecting harnesses.

To combine the science functions into a single package and facilitate an 
easier assembly, integration and verification program, a modular optical 
design was chosen where lower level assemblies could be manufactured and 
their performance verified prior to being brought together in the complete 
instrument. This approach also enables parallelism and flexibility in the 
build, test and qualification flow but places stringent requirements on the 
`systems engineering' component of the project, with interfaces between 
sub-systems needing to be defined, controlled and monitored carefully at all 
stages. The design solution resulted in the OM being split among four main 
optical modules (subsystems), as shown in Figure 2a and listed as follows,

\begin{itemize}
\item Interface Optics and Calibration (IOC)
\item Mid-infrared imager (MIRIM), interfacing to one Focal Plane Module (FPM) with its detector array. MIRIM encompasses the imager, low resolution spectroscopy and coronagraph modes of the instrument.
\item Spectrometer Pre-Optics (SPO)
\item Spectrometer Main Optics (SMO) interfaced to two FPMs with detector arrays; the SPO and SMO constitute the Medium Resolution Spectrometer (MRS).
\end{itemize}

These modules were integrated onto a single structure, the Deck. The 
completed OM is mounted to the JWST ISIM via a carbon fibre reinforced 
polymer (CFRP) hexapod mounting system (the black rods in 
Figure 1). This hexapod thermally isolates the OM 
from ISIM, which is passively cooled to about 40 K, while supporting it 
against the mechanical loads encountered during launch \citep{jessen2004}.

The MIRI optics take full advantage of state-of-the-art large-format 
mid-infrared detector arrays. Three focal plane modules (FPMs) with 1024 X 
1024 pixel Si:As IBC detector arrays (Rieke et al. 2015b, Paper VII and Ressler et al. 2015, Paper VIII) interface to the 
OM, with one array dedicated to imaging, coronagraphy, and low resolution 
spectroscopy, and the other two used in the medium resolution spectrometer. 
The FPMs attach to the outside of the optics modules, mating two flat 
surfaces (with locating fixtures) to provide robust and accurate alignment 
onto the outputs of the instrument optics.

\subsection{Thermal and Cryogenic Considerations}~

MIRI is the only instrument that must be cooled below the temperature
achieved by passive cooling of the ISIM to optimise the detector
performance and reduce the thermal background below the detector dark
current. The design was developed with thermal constraints as a key
driver. The optics module is maintained at a temperature below 7 K by
the cooler system.  Because of its well understood structural and
thermal behaviour, aluminium alloy was used to make the supporting
structure of the deck and the four optical subsystem modules. The reflective optical
surfaces are also of aluminium to simplify the thermo-mechanical
design and for the stability of alignment during cool down. This
approach had been proven for other flight and ground based instruments
in the mid and far-IR (e.g., IRS and MIPS on Spitzer, SPIRE and
PACS on Herschel, VISIR on ESO-VLT, Michelle on Gemini/UKIRT) but has
been taken to higher levels of precision in MIRI. The instrument
optical subsystems and the FPMs are built and aligned at room
temperature, and remain aligned when cooled. The designs of both the
imaging and spectrometer channels were implemented using the minimum
number of low power cryogenic mechanisms (section 6) to minimise the
heat load to the cooler.

\section{Optical Design}
\label{sec:optical}
The optical paths through the instrument are shown schematically in Figure 2b. 
Both the Imager and Spectrometer channels are fed from a single pick-off 
mirror in the IOC. The region of the focal plane intended for MIRIM is then 
selected by a fold mirror close to the telescope focal plane, with light 
intended for the MRS allowed to pass on through the deck. The positions of 
the fields in the V2, V3 coordinate system relative to the JWST telescope 
boresight at V2 $=$ V3 $=$ 0 are shown in Figure 3.

\subsection{Imager, Coronagraphs, and Spectrometers}

Inside MIRIM, the field of view (FOV) is partitioned into three functional 
areas; imager, coronagraph and low-resolution spectrometer as indicated in 
Figure 3, enabling all science functions to be supported by a single 
detector array and a single wheel mechanism. The light is collimated and, at 
the pupil image formed by the collimator, the single wheel holds the filters 
for the imager and coronagraphs, a prism assembly for the low resolution 
spectrometer, a blank for dark current measurements and a pupil imaging 
lens. This entrance focal plane is imaged onto the detector using a 3-mirror 
anastigmat camera with separate areas of the detector being dedicated to the 
imaging, coronagraphy and spectroscopy functions. A full description of 
MIRIM can be found in Paper III. 

The MRS (Paper VI) provides diffraction limited integral field spectroscopy 
over the whole wavelength range from 5 to 28.5 $\mu $m. It consists of two 
modules shown in Figure 2b -- the SPO, which splits the incoming light both 
spatially, to form an entrance slit for the grating spectrometer, and 
spectrally into 4 channels, each of 3 sub-bands, that are dispersed and imaged onto the detectors 
in the SMO. The wavelength range is divided into the 4 channels using dichroic 
mirrors in the SPO; the channels have separate dedicated integral field 
units and the spectra from each of the 4 channels occupy half of one of the 
2 MRS detectors. Each channel is split, by a dichroic chain, into 3 sub-bands that are observed 
sequentially by rotation of just two mechanisms that carry both the 
wavelength sorting dichroics and the dispersion gratings in a very compact 
and efficient configuration. 

\subsection{Design considerations}

The limited space allocated to MIRI, plus the need to keep the instrument 
overall as compact as possible to minimise the radiative heat load on the 
outer envelope, resulted in the use of relatively fast optical beams. These 
optics are designed to operate without vignetting and to meet image 
quality requirements in the presence of up to 4\% pupil shear (i.e., the mis-alignment 
of the telescope exit pupil and the instrument entrance pupil in units of their diameter) and 
2mm of focus offset, tolerances that became requirements for the optical 
alignment strategy.

A tolerance analysis showed that MIRI would not need a focus
mechanism, so long as tight alignment tolerances were maintained to
place the focus position onto the detector and to position MIRI onto the
telescope. The design solution to ensure the detector surface was
placed correctly was to measure both sides of the flat interface plane
and then machine a dedicated ``shim surface''. The detector surface
position was measured both warm and at cryogenic temperatures to
support this approach. To place the output of the optics correctly, an
alignment budget was created that gave pupil shear and focus
allocations to the sub-systems (IOC, MIRIM, SPO, SMO); to each of the
interfaces between these sub-systems and the deck; and to the
ISIM-MIRI hexapod mount. These budgets were set with the intention of
achieving an overall focus within 1mm and pupil shear of no more than
2\% for MIRI.  The imager and spectrometer were also required to be
confocal, which was achieved via the mechanical alignment of the
subsystems to the deck. The all aluminium structure means that all
sub-system interfaces are direct between mounting pads on each
surface, fixing 1 lateral and 2 angular degrees of freedom, with
dowels defining the other 2 lateral and 1 angular degrees of
freedom. The tolerances on these mechanical interfaces and Monte Carlo
analysis showed that alignment within budget was possible without
recourse to a measure-adjust-measure cycle.

\subsection{Alignment into JWST}

The optical alignment of MIRI with respect to the telescope requires
that the position and orientation of the entrance focal plane and the
entrance pupil coincide with the focal surface and exit pupil of the
telescope, respectively. Achieving this accurately is key to the
scientific performance of the instrument. The positions of the
telescope focal surface and exit pupil are well-defined with respect
to the telescope optical elements and the mechanical interface between
MIRI and the ISIM. However, these definitions are for the in-orbit
environment of cryogenic temperature and zero gravity.  The design of
MIRI needed to take into account the offsets that will occur to the
system from the warm as-built conditions found in a terrestrial lab.

On cooling, the distances between the CFRP leg to deck mounting points 
decrease by the integrated CTE (coefficient of thermal expansion) of 
aluminium, and the legs shorten by the integrated CTE of CFRP. Analysis 
showed that the leg/deck interface points would move towards the MIRI/ISIM 
interface by 0.49 mm on cooling and by only $\sim$ 20 microns when 
gravity is reduced to zero. At the same time, cooling of the optical bench 
causes the pickoff mirror (POM) to move towards the leg/deck interface. 

The optical design model of MIRI was used to find by analysis the warm position and 
orientation of the POM that simultaneously placed the telescope focal plane 
at the MIRI entrance focal plane and the MIRI entrance pupil and telescope 
exit pupil at the same location when the system is cooled to its operating 
temperature. This warm position of the POM was used to inform the design of 
the IOC and to define the nominal warm positions of the MIRI entrance focal 
plane and entrance pupil. These were used for alignment verification during 
the room temperature construction of the optical subsystems and their 
integration into the OM. 

Prior to delivery to NASA the overall alignment of MIRI was checked at
room temperature using a NASA supplied reference system called the
ASMIF which reproduced both the mechanical and optical interfaces
within the ISIM and hence ultimately to the telescope. A series of
measurements of pupil shear and focus were made, before and after
vibration and cooling to operating temperature, using the references
built into MIRIM.  The data show that, with measurement uncertainties
of 0.35\%, the MIRI contribution to pupil shear is 1\% and there is no
discernable change with changing gravity vector. Focus measurements
demonstrated that the MIRI focus is within 0.5 mm of the nominal
position. The relative alignment between the entrance pupils of the
MRS and MIRIM was measured at cryogenic operating temperature by
scanning a point source across one quadrant of the MIRI pupils and
correlating the resulting pupil maps to the as-built optical design models of
the MRS and MIRIM. No measureable offset between the imager and
spectrometer pupils was found. The warm pupil measurements were
repeated after delivery using the same fixture to verify that there
had been no unexpected issues arising from the transfer.  Pupil shear
and focus of MIRI relative to the nominal position within ISIM have
subsequently been measured at NASA Goddard at cryogenic operating
temperature, confirming the warm measurements. All of these results
comfortably meet the targets set in the alignment budgets (\S3.2) to
have no significant impact on the performance of MIRI.

The excellent alignment of MIRI determined during 
test, and the end-end performance discussed in 
Papers III, IV, V, and VI demonstrate the success of the opto-mechanical 
approach to the MIRI optical design and alignment.

\subsection{Stray light control}

Careful attention has been paid to stray light control. The fine steering 
mirror (FSM) within the telescope optics is surrounded by a cold stop that 
provides the defining cold baffle around the primary mirror. Cold pupil 
stops are provided within each of instrument modules. They are slightly 
over-sized to avoid vignetting at the FSM stop even in the presence of a 
small level of pupil shear, so they provide an additional level of stray 
light rejection without affecting the optical path. Papers III and VI 
describe the straylight suppression features within the MIRIM and the MRS. 

\subsection{On-Board Calibration}
Stable sources of illumination are needed on-board MIRI for calibration of 
the instrument's response close in time to an astronomical observation. The 
requirement is to achieve high signal to noise in a short exposure time to 
derive high spatial frequency flat fields (pixel-pixel gain matrix) and for 
the source to be sufficiently stable that it can be used to monitor relative 
detector gain between observations of standard stars. The illumination 
should therefore be smooth on a spatial scale larger than one pixel and 
stable for timescales of tens of days. 

Identical calibration sources are provided for both the imager and the 
spectrometer (Glasse et al. 2006). One source is mounted in the spectrometer 
pre-optics and light is injected into the spectrometer optics via a hole in 
a folding flat mirror. For the imager the source is mounted in the IOC and 
light injected via a small relay mirror. 

Figure 4 shows the source design. Pseudo blackbody radiation is produced by 
miniature tungsten filament lamps and is rendered uniform by a diffusing 
surface within an integrating sphere. There are two filaments in each sphere 
for redundancy. To avoid the steep fall-off at short wavelengths in the 
blackbody spectrum, the filament must emit with an effective temperature of 
at least 500 K. In practice the operating temperature is restricted to less 
than 1000 K to maximise the filament lifetime. 

The long-term stability of the MIRI internal calibration sources was 
verified during the Flight Model (FM) campaign. The relative flux and 
repeatability of the source current were measured on eight occasions spread 
over a period of 46 days. Prior to each measurement the detector was 
annealed to ensure the results were not affected by its previous history 
(e.g. latents, image persistency or other detector issues, Ressler et al 
2015, Paper VIII). Signals through the flight filters called F560W, F1130W, F1800W, and F2100W 
(respectively at 5.6, 11.3, 18, and 21 $\mu $m, Bouchet et al 2015, Paper III) were 
measured, to provide a good representation of the wavelength range covered by the MIRI imager. 

The relative flux from the source was defined as
\[
\phi \left( filter,t \right)=\frac{\varphi (filter,t)}{\frac{1}{n}\times 
\sum\nolimits_{t=0}^n {\varphi (filter,t)} }\times \thinspace 100
\]
where t is the measurement number, and $\varphi $ is the measured average 
flux in DN/s within a single photometry region (three boxes of 100x100 
pixels$^{\mathrm{2}}$ and one of 200x200 pixels$^{\mathrm{2\thinspace 
}}$using clean areas of the detector). The standard deviation of the 
relative flux value over time ranged from 0.411{\%} to 0.270 {\%} depending 
on the filter. These values were afterwards corrected to account for 
variations in the MIRI Instrument Control Electronics (ICE) calibration 
source drive current, where the corrections were derived from the accurately 
sampled current values recorded in telemetry. This correction has been 
implemented in the Flight Software as an autonomous adjustment, to be 
applied to the source current once after every switch on. 

The final, corrected, calibration source relative flux stability was found 
to range from 0.039 {\%} to 0.203 {\%} on a per filter basis, over a 46 day 
period. This compares very favourably with the accuracy of absolute flux 
calibration using standard stars which is estimated to be about 1 {\%}. 

\section{Mechanical Design}

\subsection{Mechanical Configuration and Requirements}

The MIRI OM is of isothermal construction with the Deck 
and the optical subsystems all constructed in aluminium alloy and thermally coupled 
together by bolted interfaces. The combination of this all-aluminium assembly with
a simple and efficient CFRP hexapod provided a well understood structure, which
could be specified and built to warm dimensions and whose offsets at
cryogenic temperatures and under zero-gravity could be accurately predicted.

The main driving requirements for the 
structure as defined at the outset of the programme are listed in Table 1. 
Sizing of the Deck and the Hexapod elements was driven by these requirements.

The deck is a ribbed and pocketed structure designed to support the various elements with the 
least expenditure of mass, maximum stiffness, maximum stability and lowest technological risk.  
It is machined in four parts from aluminium alloy 6061. The support for the spectrometer is made as a 
single part, carrying interfaces for the two Spectrometer Main Optics modules, the Spectrometer Pre Optics and 
the Hexapod.  It is bolted onto the ''lower deck'' structure that carries interfaces for the 
Cooler 6K Heat Exchanger, 
the Imager and the Input Optics and Calibration assembly plus some ancillary items. Two struts bolted to this 
lower deck help to support the wider side extensions to the spectrometer part. The Deck is highly 
lightweighted and the pocketing is such that there is significant mass only in the regions where 
the subsystem interface pads locate.
The Deck provides the stiffness between the Hexapod apices and was qualified early in the programme 
using the Structural Thermal Model (STM). 

\subsection{Hexapod Design and Test}
The hexapod struts were manufactured from Carbon Fibre Reinforced Plastic 
(CFRP), which has a favourable combination of strength and low thermal 
conductivity at cryogenic temperatures. In practice, the hexapod design is 
stiffness limited, with the driving goal being to minimise the thermal cross 
section whilst maintaining a margin on the first frequency requirement. The
second important design driver for the hexapod struts is buckling. High stiffness and high buckling resistance 
is achieved by having a high Young's modulus in the axial direction of the strut. 
This implies a lower cross section to achieve the first eigenfrequency, a key requirement in 
Table 1. The chosen design solution for the hexapod is six struts of length 405mm, 
diameter 35.5mm and 1.2mm wall thickness. This  
sizing avoids buckling under the design load and maintains a damage-tolerant wall thickness \citep{jessen2004}. 

The characteristics of the (T300) fibre used in the hexapod are well 
documented and their use has significant space heritage. This high strength 
fibre was preferred over a high modulus fibre due to its lower sensitivity 
to micro-cracking when cycled to cryogenic temperatures. The resin system is 
L20/SG. The hexapod strut end fittings and brackets are made from invar for 
thermo-elastic compatibility with the CFRP.

The Hexapod struts went through an extensive test campaign at strut level to qualify the manufacturing process and confirm strut performance.  The performance of the Hexapod and Deck to provide the required alignment stability and withstand expected launch loads was verified by the Flight Model test campaign. 

\subsection{Assembly and Alignment}

As described in sections 3.2 and 3.3 the internal alignment of MIRI and the alignment into JWST 
are achieved by mechanical design and tolerances.  To meet the required alignment performance
the majority of the structural parts and the mirrors are constructed of two compositionally 
similar, heat treatable aluminium alloys, 6061 and 6082. Structural and 
optical components were thermally aged to ensure adequate dimensional 
stability through subsequent temperature cycling. The optical subsystem modules were 
mounted on the Deck using bolted and dowel-pinned interfaces. The Deck was machined to a surface flatness 
of 20$\mu$m throughout and the dowel hole location tolerances ranged from 25$\mu$m for the IOC to Deck interface to 90$\mu$m for the SMO to Deck interface. We note that 
no optical misalignments have been seen during testing of the MIRI Flight 
Model. 

Repeatable mechanical location of MIRI with respect to test and flight 
interfaces was achieved by means of 6 quarter inch dowel pins, 2 at each 
hexapod `foot'. For the flight interface to ISIM, these pins locate in a 
hole and a slot per foot on the ISIM side of the interface. 

Conventional practices dictate that ground support equipment to handle the instrument should not occupy 
flight interfaces.  However, transportation (which is the most severe environment seen) was carried out 
using appropriately protected flight interfaces.  This was because of the need to control mass and 
the presence of an assembled, alignment critical and unconstrained friction locked hexapod. (For all other 
purposes there are lifting brackets on the Deck conveniently close to the centre of mass). To avoid 
accidental damage to the hexapods, a system of `tie rods' was employed to 
support their feet when the instrument was not mounted at its mechanical 
interface. The tie rods incorporate length adjustment such that the foot 
positions could be micro-adjusted to accommodate manufacturing tolerance 
differences among the various test, flight and transport equipment 
interfaces that the instrument would be mounted to.  This system resulted in the 
overall pupil shear and focus measurements reported in section 3.3.

\subsection{Mechanical Loads}
The sine loads dictated the design case for the semi-kinematic MIRI OM, as 
there are structure resonances in the range 50 to 100 Hz. Notching of these 
input loads during test was essential to protect the flexures in the end fittings
from damage. The vibration test approach and results are described in detail in Sykes et al.,2012.

The random vibration levels specified at the MIRI instrument interface are 
relatively low, but nevertheless, attention must be paid to the critical 
subassemblies during test to ensure that subassembly specifications are 
not locally exceeded during instrument test. In particular, the mechanism 
vibration tests were heavily notched. The notched subassembly test inputs 
became constraints on the instrument level test as secondary notch limits. 

As a basis for interface design and primary notching, limit loads were 
specified. The MIRI vibration test was a force limited test, meaning that 
acceleration input was controlled such that the measured force at the 
interface would not exceed a predetermined maximum. The maximum was set by 
direct reference to the design load for sine vibration, or by reference to 
the NASA semi-empirical method (Scharton 1997), in the case of random 
vibration. By this method the launch loads were verified to be enveloped 
without the excessive over testing that would result from applying nominal 
vibration inputs through the main structural resonances.

\section{Thermal Design}
\label{sec:thermal}
\subsection{Overview}
MIRI is the coldest instrument on the observatory. The detectors themselves 
(Papers VII and VIII) must be held at a nominal temperature of 6.7 K with a 
temperature stability range of 20mK over a 1000 second exposure. The deck 
and optics must be held at a temperature below 15.5 K with a stability 
within a 1K band to avoid background radiation at long wavelengths that 
would impact the system sensitivity. It was decided to cool the whole deck 
and modules attached to it to $\sim$ 7K, near the detector 
temperature, to remove temperature gradients and therefore possible sources 
of mis-alignment on cooling. 

The total heat load from the OM to the cooler heat-exchanger stage during 
nominal operation must not exceed 46.5 mW. The nominal time-averaged 
dissipations internal to the OM are determined to be 10.46 mW by correlation 
of measurements with the MIRI thermal and operating models, leaving about 36 
mW maximum for conductive and radiative loads. 

To isolate the OM from heat generated in the ISIM that might undermine its 
thermal design, the OM is enclosed within a cooled shield at a temperature 
of around 23 K (Figure 5). The OM is conductively isolated from the ISIM by 
the hexapod struts visible in Figure 1, which are 
attached to the ISIM conductive interface, having a temperature of about 40 
K. The OM and shield are cooled actively by couplings to the 6 K stage and 
the 23 K Heat exchanger stage assembly of the MIRI-dedicated cryo-cooler, 
which is described below. 

\subsection{Cooler}

The cooling to the detectors, OM, and thermal shield is provided by a 
$\sim$ 6 K/18 K hybrid mechanical cooler, provided by Northrop 
Grumman Aerospace Systems in collaboration with JPL. The system is a further 
development from the NASA Advanced Cryocooler Technology Development Program 
(ACTDP), which achieved a breakthrough in cooler efficiency while achieving 
heat lifts of 30 mW from 6K and 150 mW from 18K (Ross 2004). 

The MIRI Cooler System uses helium as the working fluid and consists of a 
three stage Pulse-Tube (PT) Pre-cooler that reaches $\sim$18 K and 
a fourth $\sim$6 K stage, which is a Joule-Thomson (JT) cooler\footnote{A 
JT cooler works on the familiar principle of allowing compressed gas to 
expand as it passes through an orifice. Pulse tube coolers (e.g., Radebaugh 
2000) produce an oscillating flow through an orifice, or more commonly 
through a thermal matrix called a regenerator. In the high pressure part of 
the cycle, warm gas is driven into a reservoir, where it exchanges its heat. 
In the low pressure part, the gas flows back through the regenerator and 
cools it, allowing heat to be removed from the object being cooled.}. 
These two temperatures are made available to the instrument through heat 
exchangers. It is characteristic of JT devices that their heat lift 
decreases substantially with increasing precooling temperature. Therefore, a 
valve bypasses the JT expander to allow the initial cooldown of the 
instrument by the pulse tube stages. At $\sim$ 18K, the bypass 
valve is closed and cooling by the JT expander continues to 6K. This 
crossover is termed the ``pinch-point" because it is the temperature where 
there is a minimum in overall heat lift capability. 

The cooler system architecture is made particularly challenging since the 
cooler spans the length of the JWST observatory (Figure 6). The Cooler 
Compressor Assembly (CCA) is in the JWST Spacecraft Bus and is at room 
temperature, while the Cold Head Assembly (CHA) is mounted 
on the ISIM structure near the MIRI OM. Both the compressors (PT and JT) are 
driven by the block-redundant Cooler Control Electronics Assembly (CCEA) 
which is located in JWST Spacecraft Bus. 

The CCA consists of the PT-Pre-cooler, the PT and JT Compressors (see Figure 
7), plus a radiation shield for the various stages of the pre-cooler. It 
also supports the stowage structure for the Refrigerant Line Deployable 
Assembly (RLDA) that runs through the observatory to carry the working fluid 
from the CCA to the CHA and back. Finally, it includes the structural 
element that mounts to the spacecraft bus for launch and provides thermal 
interface to the JWST heat rejection system. The refrigerant lines are 
supported by thermally isolating line supports. The CCEA provides the drive 
to the compressors and also implements the functions of thermal control, 
compressor vibration reduction, telemetry generation, and heater and valve 
control. The CHA contains the Joule Thomson constriction and has the 
cryogenic valve that affords switching between the Pre-cooler mode and the 
JT-Cooling mode during the MIRI optical system cool-down. Another valve 
bypasses more of the cold assembly to allow warming the cooler lines for 
decontamination.

A flight-like CHA was delivered in the spring of 2013. It, along with a 
ground support equipment pre-cooler, successfully supported the cool-down 
and operations of the MIRI OS during the ISIM Cryovac1 Test (CV1-RR). The 
flight CCA and the CCEA are currently in development.

The MIRI cooler system operations will be verified and validated through a 
series of acceptance test programs followed by a MIRI End-to-end test where 
a full complement of flight-like cooler hardware will be tested along with 
thermally representative MIRI hardware (the STM, summarised in section 1). 
The End-to-end test will include a flight-like MIRI Thermal Shield, which is 
also cooled by the MIRI pre-cooler.

\subsection{Conductive isolation}

The CFRP hexapod is one of two main conductive paths between the MIRI OM and the 
ISIM. The conductance of each strut is about 0.02 mW/K at a mean temperature 
of 10 K increasing to about 0.06 mW/K at a mean temperature of 40 K 
(Shaughnessy et al. 2007). Table 2 summarises the main conductive and radiative heatloads between the MIRI OM and the ISIM or Shield calculated from the correlated thermal model.

Electronic harnesses and a purge pipe also provide paths for conductive heat 
loads. Harness loads are managed by: (1) minimizing of the number of wires 
required, (2) definition and control of the effective thermal length of the 
harness between the ISIM heat-sink and the OM deck, and (3) selection of low 
thermal conductivity materials for construction of the harness (manganin, 
phosphor bronze and stainless steel). The purge pipe (which is primarily 
needed to maintain a clean and dry environment for the wheel lubricants to 
allow their operation during non-vacuum test activities), is constructed 
from stainless steel, sized to minimize the conducted load, whilst allowing 
the required volumetric flow rate to provide a suitable positive pressure in 
the instrument.  

\subsection{Radiative isolation}

With the exception of the optical aperture and access for the cooler heat 
exchanger, the outer surface of the OM is covered with Single and 
Multi-Layer Insulation (SLI and MLI) blankets. The SLI is used around the 
six struts to minimize the conducted heat load from the warmer ISIM. The SLI 
also encloses harnesses and the purge-pipe that are attached to the struts. 

There are two issues regarding MLI for MIRI. First, at low
temperatures the thermal isolation achievable is small (e.g., Spradley
et al. 1990).  Nonetheless, it was decided to encase the OM in MLI
both for the thermal gain and as a protective measure for its
low-emissivity surfaces. Second, the skin depth of aluminium at a
wavelength of 400 $\mu $m (by Wien's Displacement Law, corresponding
to a temperature of $\sim$ 7K) is about 0.1 $\mu
$m. Therefore, the 1000 {\AA} (0.1 $\mu $m) aluminium coating on
typical blanket materials can be somewhat transmissive to thermal
radiation originating from MIRI and the cold ISIM
environment. Consequently the MLI blankets were constructed using
Kapton having a 5000 {\AA} (0.5 $\mu$m) deposition of aluminium on one side. Eight
layers are inter-leaved with crinkled double aluminized Mylar to
inhibit conductive heat transfer (net spacers are not used as they are
a source of particulate contamination).

\subsection{Thermal control}
The detectors are thermally isolated within the FPMs and cooled using a 
thermal strap of high purity copper from the mount for the detector arrays 
to a thermal connector on the FPM housings. From this connector, another 
strap runs approximately 0.5 m from the FPMs to the interface point near the 
6 K heat exchanger. This section of strap is constructed from two 2 mm 
diameter pure (99.999{\%}) aluminium wires that are clamped into specially  
designed end-fittings to maximize interface conductance. The FPM strap has 
been sized to allow heating the detectors to keep them warmer than the OM 
during cool-down (for contamination control), and to permit annealing cycles 
where the detector temperature is raised briefly to $\sim$ 15 K. 

\subsection{Contamination Control}

As a consequence of the MIRI being colder than the rest of JWST, particular 
attention was paid in the design to contamination control. There is natural 
protection from the thermal isolation until the MIRI cooler is activated. 
Once that occurs, the instrument cools below its surroundings and 
contaminants from them can collect. Models of the expected JWST outgassing 
indicate that in the worst case, without protection, a 1.5$\mu $m layer of 
water ice along with various organic contaminants that are still volatile at 
40K could accumulate on exposed MIRI optical surfaces that are at 7K. These 
considerations led to a design with a contamination control cover (CCC) just 
inside the optical train after the pick-off mirror, which is thermally 
isolated from the rest of the optics and can be decontaminated by heating. 
The first cold surface is most vulnerable; the long path of the IOC protects 
optics further down from transported contaminants when the CCC is open. The 
CCC design is discussed in section 6. 

The Pick-Off Mirror (POM) is the coldest exposed optical surface within the 
ISIM. It is thermally isolated within the OM structure and fitted with 
redundant heaters to allow it to be warmed if necessary to drive off 
contamination (solid N$_2$, O$_2$, H$_2$O, CO$_2$) that may stick to the mirror. Before 
the POM heater is activated the CCC is closed to ensure contaminants do not 
freeze out onto the sensitive internal surfaces.

\subsection{Thermal Model Verification and On-Orbit Prediction}

A cryogenic test facility was developed at the Science and Technology 
Research Council's Rutherford Appleton Laboratory to simulate the 
environment of the ISIM, including the 40 K radiative environment specified 
at that time (i.e excluding the Thermal Shield described above) and the 
conductive interface with the OM and the 6 K heat-exchanger (Shaughnessy \& Eccleston 
2009). A three-month cryogenic test was undertaken on the Flight Model 
OM to verify and calibrate its performance and to assess the thermal 
sub-system. Two phases of dedicated thermal tests provided steady-state and 
transient data for validating thermal models. A close correlation of heat 
load and temperatures was made to the steady-state data. Heat loads were 
correlated to within 0.5 mW of measurements and temperatures were correlated 
to well within 100 mK of measurements. 

The nominal steady-state heat load predicted with the correlated model is 
33.8 $\pm$ 6 mW. This shows a margin of 6.7 mW from the requirement of 
46.5 mW, demonstrating that the cooler sub-system will be able to cool and 
maintain the OM at the required operating temperature.

The in-orbit cool-down prediction using this correlated model is presented 
in Figure 8. In the model, the ISIM boundary temperatures follow specified 
profiles, shown in the figure. The temperature of the MIRI shield shown is 
also interface data for the cooldown prediction and includes the response of the shield to the 
pre-cooler. The OM is cooled passively until it reaches approximately 100 K, at which point
the cooler is activated.  A temperature-dependent 
cooler heat lift was 
provided for analysis. To demonstrate margin on the requirement, the effective heat 
lift in the model was reduced by 25{\%}.

The analysis predicts that it takes about 110 days for the OM to reach 
operational temperatures. The cooler is activated after about 80 days. The 
inflection in the OM temperature just past 100 days marks the transition 
through the pinch-point. For contamination control, the OM critical optical 
elements are required to remain above 165 K until the ISIM is 140 K or 
below. The analysis confirmed that the OM cool-down lags that of the ISIM 
and cools below 165 K about 15 days after the ISIM passes 140 K. 

\section{Mechanism Design}

The MIRI Optical System contains four cryo-mechanisms: (1) an 18 position 
filter wheel assembly (FWA), mounted in the imager (see Figure 9); (2) two 
combined grating/dichroic wheels (DGA-A and DGA-B) with three positions each 
(see Figure 10) in the spectrometer; and (3) the 
contamination control cover (CCC) at the entrance of the optical path of the 
instrument (Figure 11). 

The wheel and grating mechanisms lie at the heart of MIRI science 
operations. The filter wheel assembly is required to achieve a high 
positional accuracy and repeatability to enable precise alignment of the 
coronagraph pupil stop. The tight repeatability requirement for the 
dichroic-grating wheels is derived from the need to move a wheel to select 
a new wavelength range without recalibrating the wavelength scale. 

The wheel and grating mechanisms are all based on the same principle: the 
wheel bodies are pivoted in a central combined bearing and retained in their 
optical position by a ratchet system. A brushless (and gearless) central 
torque motor is used to operate the wheel in an open loop drive. This 
requires only relatively simple but robust drive electronics. In addition it 
minimizes the number of harnesses from the warm electronics to the cryogenic 
part of the instrument and thus the conducted heat load. The chosen wheel 
design guarantees high precision and highly reliable positioning of the 
optical elements while using low driving power -- in particular zero power 
during science operation -- and therefore low heat injection into the cooled 
MIRI instrument (more details can be found in Krause et al. 2010). 

Operating the wheels from one position to the next adjacent position takes 
$\sim$ 8 seconds in total. This includes $\sim$ 500 
milli-seconds for motor acceleration and deceleration, $\sim$ 3 
seconds of settling by the ratchet system and $\sim$ 4 seconds to 
complete a final position sensor readout to crosscheck that the correct 
position has been reached. The final positioning accuracies are 
$\sim$ 1 arcsec for the FWA and $\sim$ 3 arcsec for the 
DGAs. 

Since no wheel angle feedback is available during the movement, the precise 
characterization of the mechanisms and their motors was fundamental to 
minimize heat load and maximize the reliability of the mechanism movements 
over their lifetime (Detre et al. 2012). This has been achieved and proven 
over several test campaigns.

The Contamination Control Cover (CCC, Glauser et al. 2008) is a door 
mechanism located in the Input Optics between the MIRI pick-off mirror and 
the first fold mirrors (see Figure 11). The CCC was introduced to protect 
the instrument against molecular contaminants outgassing from nearby 
structures after launch, during ISIM cool down, or during any ground based 
test campaign. With its contact-free labyrinth seal, the CCC also closes the 
instrument in an optical sense, blocking any stray-light. Since the CCC 
operates at the same temperature as the rest of the instrument, it is also 
suitable to provide a dark environment for internal calibration 
measurements. 

The CCC uses two identical redundant stepper motors that lever the cover 
towards its open position, while two redundant springs push it towards the 
closed position. The qualification of this mechanism has shown that the 
design is highly robust and reliable (Glauser et al. 2008). The molecular 
throughput has also been measured (Glauser et al. 2009) and shows perfect 
agreement with theoretical predictions. 

\section{Electronic Systems}
\label{sec:mylabel1}

The MIRI electronic systems split functionally into the electronics for the MIRI Cooler System 
that is described in section 5.2 and the MIRI Optical System electronics.

The MIRI Cooler Control Electronics Assembly (CCEA) is a set of independent and 
dedicated electronics assemblies, which control and drive the Cooler Thermal Mechanical 
Unit's (TMU) two compressor assemblies - Pulse Tube (PT) and Joule Thomson (JT). These are based on 
heritage designs currently in other space flight applications and are capable of highly 
accurate temperature control over the temperature range from 4K to 15K. The Cooler Control 
Electronics (CCE) are 
single-string, but redundant at the box level to enhance reliability and 
meet the lifetime requirement, and there is a set of primary and redundant JT 
and PT CCEs for each compressor. A third electronics assembly, the Relay Switch 
Assembly (RSA), provides the switch to allow the use of either set of cooler electronics 
to drive the single TMU assembly. The RSA contains latching relays and accepts a pulse 
command from the spacecraft to effect switching from primary to redundant CCEs, or visa 
versa. One key function of each JT and PT CCE assembly is to convert, condition, switch, and 
distribute incoming SC primary bus power, and furnish it in the correct form to drive the 
various elements of the compressor assembly. Each CCE provides closed-loop control of 
various compressor and cold head functions, monitors the status of key performance and 
safety parameters, and communicates with the ISIM Command \& Data Handling system host 
via a MIL-STD-1553B bus. Generally these control functions involve both analog and digital 
circuitry and supporting internal software, which also provides automated fault protection.   

The MIRI Optical System electronical architecture is summarised in Figure 12. 
The operation of the instrument is controlled by the ISIM Control \& Data Handling (ICDH) system via the two 
discrete electronic boxes; the Focal Plane Electronics (FPE) and the Instrument Control Electronics (ICE). 
The Spacecraft Power Conditioning Unit (PCU) supplies power directly to each of these 
units at a nominal voltage of 31V(DC).
The FPE and ICE both operate at ambient temperatures (~300K) and are 
mounted in a dedicated warm region of the ISIM referred to as the ISIM Electronics Compartment (IEC). In addition 
to the links to the ICDH there are eight 
spacecraft temperature monitoring sensors for when the instrument is switched off.

The Focal Plane System (FPS), comprising the detector, FPE and associated harness are described in detail in 
Paper VIII and so are not discussed further here. 

The ICE controls the four mechanisms and two calibration sources discussed above, along with 
15 temperature sensors and the decontamination heater. While the mechanisms, 
sources and sensors are mounted on the optical bench at $\sim$ 7 K, 
the ICE is maintained, along with other science instrument electronic boxes, 
in a separate section of the observatory at an operating temperature of 
$\sim$ 300 K. The wiring harness connecting the OM with the ICE 
uses phosphor bronze for wires with relatively high current ($\sim$ 
100 mA), whereas stainless steel is used for the low current sensor lines. The 
ICE has no internal processing capabilities and operates only via 
command from the JWST integrated science instrument module (ISIM) control and data handling module (ICDH). 

The ICE is a fully redundant design with a modular architecture. It has two 
service modules; `DC/DC' and `TM/TC {\&} Scheduler', and two application 
modules; `Motor control' and `Conditioning.' The modules are powered, 
interconnected and communicate via a back plane, which also allows for cross 
strapping of temperature sensors and non-redundant mechanism position 
sensors, to enable monitoring with either of the two redundant sides of the 
ICE.

The DC/DC module is responsible for the power supplies handling, accommodating the spacecraft primary power 
bus (an unregulated 22 to 35V power bus).

The TM/TC {\&} Scheduler module interfaces with the ICDH via a 1553 bus and 
with the application modules via a proprietary media bus implemented on the 
back plane. As such, it receives telecommands, then translates and 
distributes these commands to the relevant application module. It also 
collects and formats telemetry data, which is then made available for the 
ICDH. Another function of this module is to drive the back plane relays to 
select the appropriate routing for the temperature and position sensors.

The Motor control module controls and monitors the drive currents and 
voltages for one (at a time) of the four MIRI mechanisms. The particular 
motor and voltage supply, (40 V nominal \textless 10 K operation, 80 V room 
temperature operation) are selected in advance via relays; then the power 
amplifier controls the coil pair for the selected motor, activating the 
mechanism. To overcome an environmental failure case of the CCC sticking in the closed 
position, a special relay setting is available to allow the activation of 
the two CCC stepper motors at the same time, thereby doubling the opening 
torque. 

The remaining control and monitoring functions of the ICE are provided by 
the Conditioning module. This module provides the current drive and 
control for the calibration sources and the decontamination heater. It 
monitors and provides telemetry data on the temperature and mechanism 
position sensors. Control of the DC/DC module, synchronisation, motor 
voltage selection and activation, is also provided in this module.

\section{Controlling MIRI}
The MIRI instrument relies on the ISIM to provide all command and control 
functions for the hardware. These services are provided by the ICDH, which consists of a single-board computer, 
some basic image processing modules, and several communications interfaces. 

The communications links between the ICDH and the MIRI electronics boxes are 
shared with the other instruments. High-speed data links are provided to the 
MIRI focal plane electronics over a Spacewire bus, routed via the ISIM 
Remote Services Unit (IRSU). The ICE and CCE (Cryo-cooler Control 
Electronics) require relatively low data rates, and are linked via the ISIM 
1553B bus. The ICDH has further links to the JWST observatory and spacecraft 
systems to allow it to receive commands, and to send science and engineering 
telemetry to the solid state recorder. Additionally the ISIM (and hence 
MIRI) receives all of its electrical power from the spacecraft. 

The ISIM flight software (FSW) consists of multiple software modules, each 
of which has a distinct function. Generic services such as communications, 
timing and memory management are provided within the `core' software, as 
these are required by all of the instruments. Each instrument has one or 
more dedicated modules for controlling its own functions, which were 
developed by the instrument teams. These modules use the services provided 
by the core ISIM software to communicate with the instrument hardware, and 
to send and receive information from the timeline or ground operator (via 
the spacecraft and observatory systems). 

The MIRI software is split into two separate modules with distinct 
functions:

1. MIRI Optical System FSW: 

\begin{itemize}
\item Command and control the MIRI OBA hardware. 
\item Operate the mechanisms, calibration sources and POM heater via the ICE. 
\item Operate the detectors and their thermal control heaters via the focal plane electronics. 
\item Monitor sensor data (e.g. temperatures) from the OBA. 
\item Maintain OBA hardware safety during commanding of each item. 
\end{itemize}

2. MIRI Cooler FSW: 

\begin{itemize}
\item Command and control the cooler system
\item Operate all cooler components via the CCE
\item Monitor sensor data (e.g. temperatures) from the cooler system
\item Maintain cooler hardware safety during operations. 
\end{itemize}

MIRI can receive commands from the ground (via the spacecraft), or from 
on-board sources such as the science timeline or stored commanding (used for 
safety-critical actions). Early in-flight operations (such as instrument 
commissioning) and some regular engineering activities will be carried out 
by ground operators, normally by using ground scripts to send commands and 
to verify the telemetry. 

Most in-flight MIRI operations will be conducted from the ISIM science 
timeline, and will be planned well in advance. The timeline will consist of 
a series of observations, each involving one or more instruments. The 
specification for each observation is translated on-board into a sequence of 
instrument and observatory operations (e.g. spacecraft pointings), while the 
MIRI commands are generated by dedicated on-board scripts for each observing 
mode. The science timeline execution system is capable of performing a 
sophisticated level of error checking, to ensure that constraints are 
enforced and that any errors reported by the instruments are handled 
appropriately. 

Commands are processed in the same way irrespective of their origin, so that 
only a single interface is needed. Most MIRI commands are reasonably high 
level, and many of them correspond to individual instrument functions. For 
example:

\begin{itemize}
\item Set cooler cold-head temperature (set-up the cooler in preparation for MIRI operations)
\item Move filter wheel (select the Imager filter required for an observation)
\item Switch on calibration source (in preparation for calibration measurements)
\item Start exposure (begin taking science data with the current detector settings)
\end{itemize}

The flight software modules are responsible for translating each command 
into appropriate instructions for the electronics boxes, and for ensuring 
that the commands are completed successfully. In addition to the high-level 
commands, there are lower-level commands to facilitate engineering 
operations and instrument troubleshooting. There are also many programmable 
options in the software that can be adjusted (e.g. operating temperature 
limits, time-outs etc.), so that any unexpected events during the mission 
can be dealt with as easily as possible. The flight software modules 
themselves can also be patched if necessary. 

The MIRI and ISIM designs lead to various constraints and limitations on 
operations. Some of the more significant examples are listed below:

\begin{itemize}
\item The three MIRI detectors can be operated individually or in parallel, but detector settings (e.g. bias voltages) cannot be altered while science exposures are in progress. 
\item The MIRI mechanisms can only be operated individually (i.e. in a serial manner).
\item Read-out of virtually all engineering telemetry from the electronics occurs at a fixed cadence of 1 reading every 4 seconds. Sampling on a finer timescale is only possible for certain mechanism parameters or in engineering modes. 
\item Science data read-out is not synchronised to engineering telemetry, and the start time of a science exposure cannot be controlled to better than the detector frame read-out time (typically about 3 s, but can range from \textasciitilde 0.1 s to \textasciitilde 27 s depending on the observing mode).
\end{itemize}

\section{The Overall Test and Verification of the MIRI Optical System}

The modular instrument design reduced risk for the flight model Assembly 
Integration and Verification (AIV) because it allowed a series of incremental 
qualification and performance verification tests to be performed at 
subsystem level. The three model approach (STM, VM, FM - structural/thermal, 
verification, and flight models respectively) proved to work well, with 
sufficient flexibility to accommodate problem solving throughout the 
programme. The main aims for the STM were to provide an early mechanical 
qualification of the Primary Structure, thermal model validation and to 
prove out the test facility prior to the VM test. The VM objective was to 
verify instrument optical performance at operating temperature sufficiently 
early to avoid major cost and schedule problems in the event of detected 
problems requiring extensive FM modifications. The VM testing was split into 
two campaigns, one to test the instrument optics with a very simple single 
point simulated JWST source and a second more extensive test using the MIRI Telescope 
Simulator, which also provided feedback to the design of the MTS and to the test plans and scripts 
for the Flight Model test campaign. 

Following integration, the MIRI Flight Model was tested for 1600 hours 
during 2011 in the test chamber described in section 5.7 and in Shaughnessy \& Eccleston 2009. 
This provided a background radiation environment that was a close analogue to the expected 
on-orbit environment, namely a near blackbody emission spectrum with an 
effective temperature of 40 K. This test campaign was therefore the best 
opportunity to measure the performance of MIRI prior to launch, especially 
in the areas of photometric calibration and straylight. Further, this test 
campaign was the only opportunity to study fully the spectral performance of 
the MRS before launch. 

\subsection{The MIRI Telescope Simulator}
\label{subsec:mylabel1}

The MIRI Telescope Simulator (MTS) was the cryogenic optical system developed to generate the 
illumination sources for MIRI performance measurements.  The MTS detailed design is described in 
Belenguer et al. (2008).  In this section we summarise its major 
functions and describe the computational model (MTSSim) that was produced to predict its photometric output.
At the heart of the MTS was a laboratory standard blackbody whose temperature could be 
selected in the range 100 K to 800 K.  The collimated beam from this source passed through an adjustable 
iris diaphragm (to set the flux level), before reaching the MTS filter wheel.  The wheel included a closed 
position 
to block the MTS hot source for background estimation; a clear position for broad band illumination; 
one long-wave pass filter and one short-wave pass filter for measurement of spectral leaks and 
four solid state Fabry-Perot etalons which provided a comb of spectral lines for the wavelength 
characterization and calibration of the MRS.  
The output beam from the filter wheel was then presented to an integrating sphere whose 
spatially uniform output was matched to the input of 
a Cassegrain telescope (the Main Optical System (MOS)).  
By inserting a pinhole  (one of two mounted on a three axis moveable stage) at this input, the MOS 
was designed to reproduce the point spread function delivered by the JWST.  The point sources could be moved to 
any point within the MIRI field of view with an absolute accuracy equivalent to 1 imager pixel, and a relative 
accuracy of better than 0.1 pixel for small displacements.  With the pinhole mechanism driven out of the beam, 
flood illumination across the full MIRI entrance focal plane was obtained. 
An infrared LED source was included at the exit pupil of the MOS which could be scanned 
across one quadrant of the MIRI pupil to measure the relative centrations of the pupils associated 
with each of the MIRI optical sub-systems.

The absolute flux calibration of the MTS was determined by modelling its optical throughput.  
This throughput estimate was embodied in a computer simulation MTSSim, a program written in IDL to calculate the 
irradiance provided by the MTS at the MIRI input plane.
MTSSim implemented a radiometric model of the MTS, with the hot source treated as a grey-body 
with an emissivity of 95{\%}. No diffraction effects were considered, and the system losses were 
limited to those caused by non-unity transmittance of the optical elements, 
as determined from sub-system measurements. Crucial to the accuracy of the end-to-end 
transmission budget was the error in estimating the transmission of the integrating sphere, since this 
could only be determined by geometrical modelling.  As discussed in Glasse et al. (2015) Paper IX, this 
uncertainty in the estimated efficiency of the MTS was regarded 
as consistent with the 55 \% difference seen when using it as a flux standard 
for measuring the throughput of MIRI, as compared with measurements of MIRI's sub-systems

\subsection{Data Analysis}

To provide a convenient reduction environment that was strictly
configuration controlled and available to all of the international
team, we developed the Data Handling and Analysis System
(DHAS)\citep{morrison2011}.

The DHAS was based on a C$++$ analysis section, with a flexible IDL user 
interface. It first converts the raw integration ramps to slopes, 
subtracting the dark signals and correcting for nonlinearity. It also 
incorporates the best known algorithms to correct non-ideal detector 
behaviour, such as the reset anomaly (commonly seen in infrared arrays; the 
first samples after a reset are offset from the rest) (see Paper VIII for 
more discussion of non-ideal array behaviour). It provides functions to 
display the resulting images and manipulate them, and also to organize the 
output of the MRS into a data cube

The DHAS essentially implements the prototype for a MIRI data reduction 
pipeline. In addition to the continued use for instrument test data (e.g., 
at ISIM level), it is also being used to test and validate algorithms for 
the more sophisticated data reduction pipeline under development at STScI. 
The DHAS is also the means by which ongoing experiments on latent images, 
subarrays, annealing optimization, and other aspects of MIRI operations are 
evaluated.

\section{Summary}
We have given a system level description of how MIRI provides its four key measurement 
functions to support a broad variety of JWST science objectives over the 5 
to 28.5 $\mu $m spectral range. Details of these functions are described in 
Bouchet et al. (2015), Kendrew et al., (2015), Boccaletti et al. (2015) and Wells et al. (2015), 
but all share a common architecture described in this paper. Opto-mechanical subsystems are mounted to an 
iso-thermal structure which is 
thermally isolated from the JWST observatory and maintained at its operating 
temperature by a dedicated cooling system. These subsystems interface with the focal planes that are 
described in Ressler et al., (2015) and Rieke et al. (2015). The MIRI Cooler, electrical 
system design, mechanisms and 
control software have been presented. We have shown how the delivered instrument has 
balanced the conflicting needs of thermal isolation against those of stiffness under the 
mechanical loads experienced during launch, low electrical power dissipation and limited mass and volume.  

The control of molecular contamination is seen to be an important consideration for an instrument 
which will be at a significantly lower temperature than the rest of the observatory.  The combination of 
a closeable cover and decontamination heaters are designed to allow scientific performance to be 
maintained throughout the JWST mission.  
The provision of on-board calibration sources complements this approach to contamination control 
by allowing all major radiometric functions of the instrument to be measured accurately and 
repeatably without recourse to any external support equipment. 

Flight Model testing of the integrated Optical System before and after 
delivery to NASA has demonstrated it to meet its key mechanical, thermal and 
optical requirements. The success of the adopted approach for achieving the 
required alignment at cryogenic temperatures by designing and testing at 
ambient and cryogenic temperatures is notable. The timely and successful delivery to NASA was 
enabled by the inherent flexibility in the programme that was provided by our coupling a modular approach to the 
build and test of subsystems with the choice of a 3 model (STM, VM, FM) system level solution for the integrated 
construction, qualification and verfication of the instrument performance.

\section{Acknowledgements}
The work presented is the effort of the entire MIRI team and the
enthusiasm within the MIRI partnership is a significant factor in its
success. MIRI draws on the scientific and technical expertise of the
following organisations: Ames Research Center, USA; Airbus Defence and
Space, UK; CEA-Irfu, Saclay, France; Centre Spatial de Li\'{e}ge,
Belgium; Consejo Superior de Investigaciones Cient\'{\i}ficas, Spain;
Carl Zeiss Optronics, Germany; Chalmers University of Technology,
Sweden; Danish Space Research Institute, Denmark; Dublin Institute for
Advanced Studies, Ireland; European Space Agency, Netherlands; ETCA,
Belgium; ETH Zurich, Switzerland; Goddard Space Flight Center, USA;
Institute d'Astrophysique Spatiale, France; Instituto Nacional de
T\'{e}cnica Aeroespacial, Spain; Institute for Astronomy, Edinburgh,
UK; Jet Propulsion Laboratory, USA; Laboratoire d'Astrophysique de
Marseille (LAM), France; Leiden University, Netherlands; Lockheed
Advanced Technology Center (USA); NOVA Opt-IR group at Dwingeloo,
Netherlands; Northrop Grumman, USA; Max-Planck Institut f\H{u}r
Astronomie (MPIA), Heidelberg, Germany; Laboratoire d'Etudes Spatiales et 
d'Instrumentation en Astrophysique (LESIA), France;
Paul Scherrer Institut, Switzerland; Raytheon Vision Systems, USA;
RUAG Aerospace, Switzerland; Rutherford Appleton Laboratory (RAL
Space), UK; Space Telescope Science Institute, USA;
Toegepast-Natuurwetenschappelijk Onderzoek (TNO-TPD), Netherlands; UK
Astronomy Technology Centre, UK; University College London, UK;
University of Amsterdam, Netherlands; University of Arizona, USA;
University of Bern, Switzerland; University of Cardiff, UK; University
of Cologne, Germany; University of Ghent; University of Groningen,
Netherlands; University of Leicester, UK; University of Leuven,
Belgium; University of Stockholm, Sweden; Utah State University, USA. 
A portion of this work was carried out at the Jet Propulsion Laboratory,
 California Institute of Technology, under a contract with the National 
Aeronautics and Space Administration.

We would like to thank the following National and International
Funding Agencies for their support of the MIRI development: NASA; ESA;
Belgian Science Policy Office; Centre Nationale D'Etudes Spatiales (CNES);
Danish National Space Centre; Deutsches Zentrum fur Luft-und Raumfahrt
(DLR); Enterprise Ireland; Ministerio De Economi{\'a} y Competividad;
Netherlands Research School for Astronomy (NOVA); Netherlands
Organisation for Scientific Research (NWO); Science and Technology Facilities
Council; Swiss Space Office; Swedish National Space Board; UK Space
Agency.

We take this opportunity to thank the ESA JWST Project team and the
NASA Goddard ISIM team for their capable technical support in the
development of MIRI, its delivery and successful integration.

We are grateful for the comments of the external referee which helped us to improve 
the clarity of high level description of the instrument in this paper.

\clearpage

\begin{deluxetable}{lc}
\tabletypesize{\footnotesize}
\tablecolumns{2}
\tablewidth{0pt}
\tablecaption{The main mechanical requirements for the MIRI structure.}
\tablehead{\colhead{Parameter}             &
	\colhead{Value}             \\
 	 }
\startdata
Initial mass budget for the OB  & 103kg \\
Minimum eigenfrequency & 50 Hz \\
Design load (including qualification margin) & 18g \\
Temperature delta across struts  &  7K - 35K \\
Maximum heat flow across struts   &  $\sim$ 6mW for 6 struts \\
Hexapod mass budget  & 5kg  \\   
Total primary structure mass budget  &  18kg \\
\enddata
\end{deluxetable}

\clearpage

\begin {deluxetable}{lc}
\tablecolumns {2}

\tablecaption{Conductive and Radiative Heatloads}

\tablehead{

\colhead {Component} & \colhead {Calculated Heatload (mW)}\\
}

\startdata

{\bf Conducted} & { }   \\
CFRP Hexapod & 7.6\\
SLI on Hexapod & 3.7\\
Purge Pipe & 0.7 \\
Harness & 5.8 \\
{\bf Radiative} & \\
ISIM to OM & 4.7 \\
Shield to OM & 0.8 \\

\enddata

\end{deluxetable}

\clearpage

\begin{figure}[htbp]
\centerline{\includegraphics[width=5.0in]{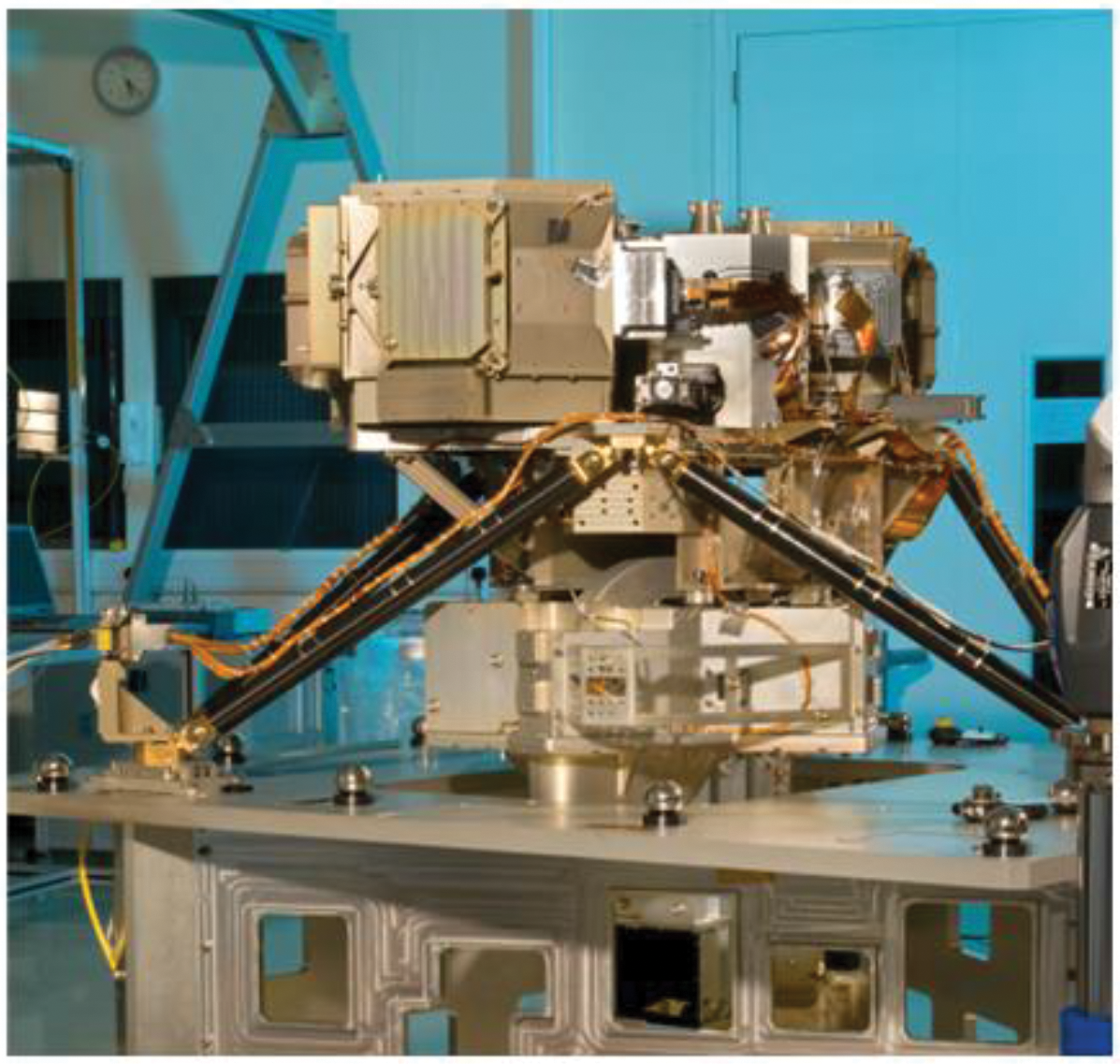}}
\caption{The MIRI Flight Model prior to delivery.  The optics module structure is aluminium, and 
it is mounted to JWST with a (black) CFRP hexapod truss..}
\label{fig:mount}
\end{figure}

\clearpage

\begin{figure}[htbp]
\centerline{\includegraphics[width=5.0in]{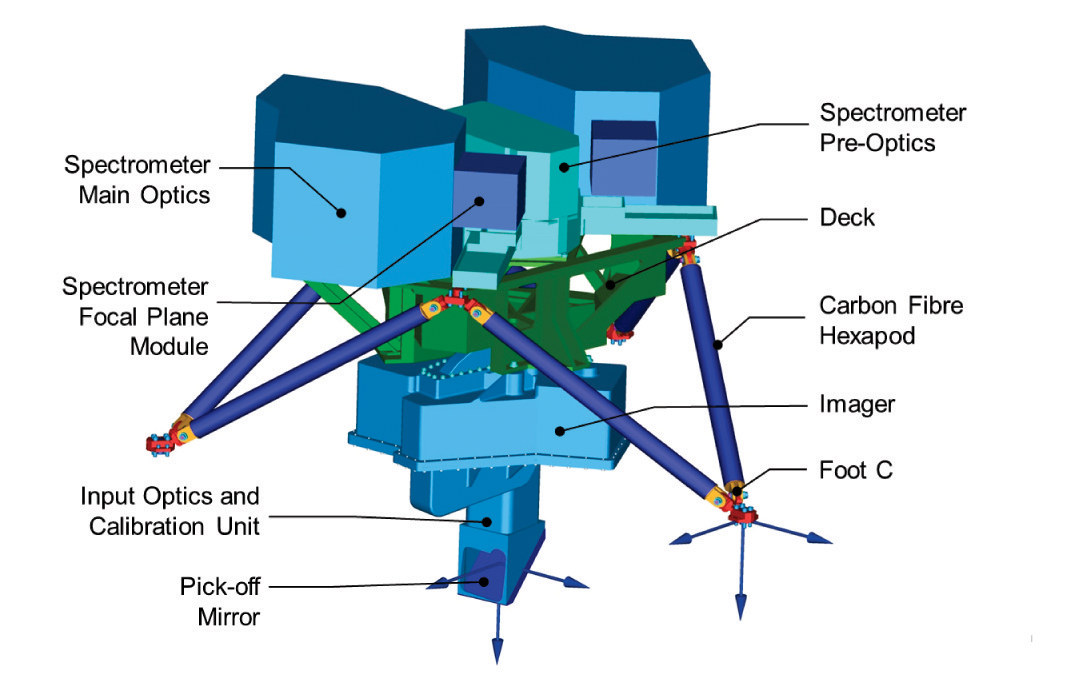}}
\centerline{\includegraphics[width=5.0in]{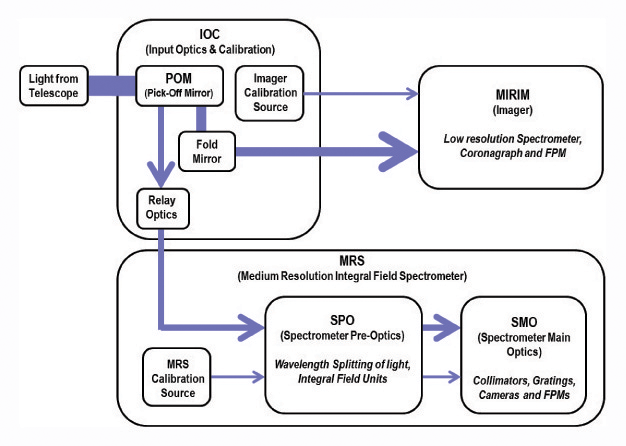}}
\caption{(upper) Overview of the MIRI optical architecture, showing the primary 
components.   (lower) The science light path (shown in blue) through the MIRI modules. }
\label{fig:mount}
\end{figure}

\clearpage

\begin{figure}[htbp]
\centerline{\includegraphics[width=5.0in]{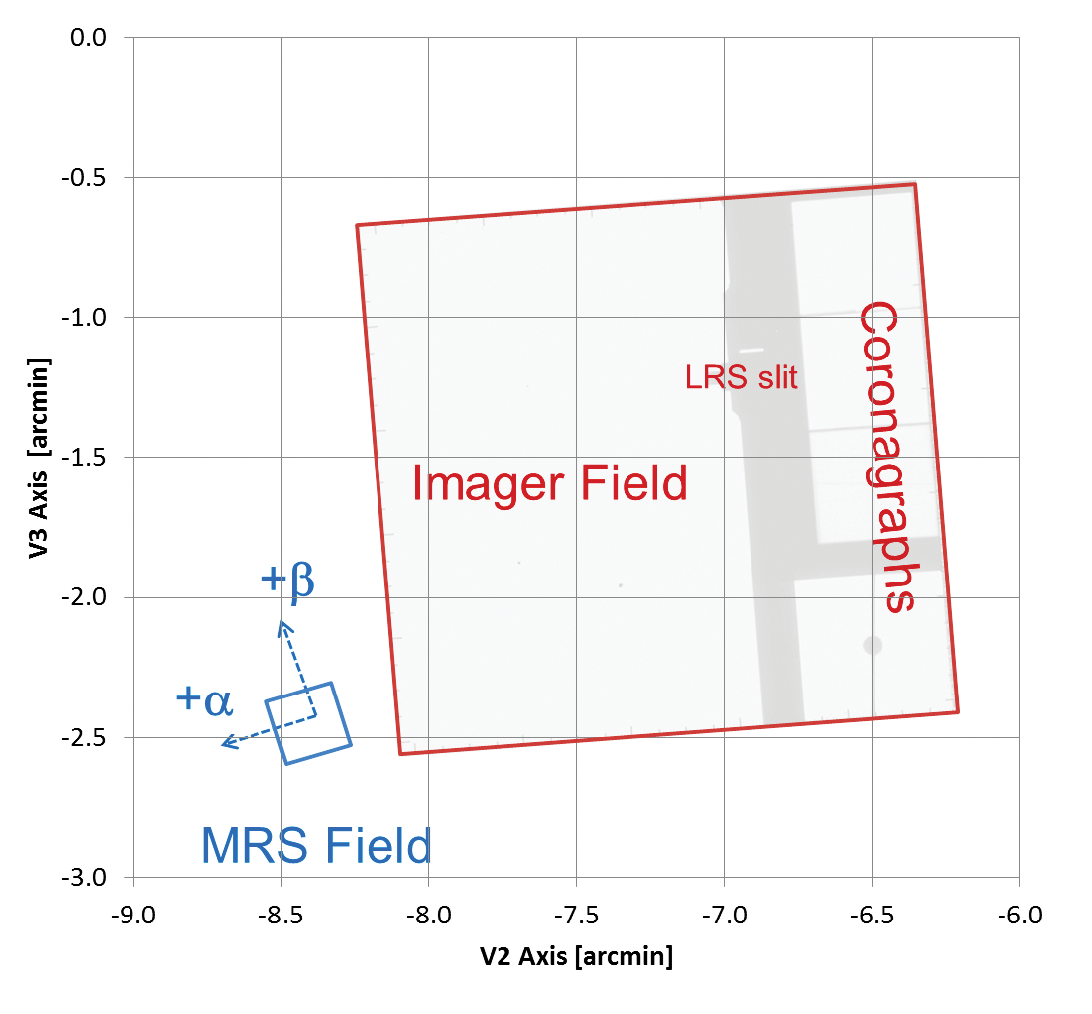}}
\caption{The positions of the MIRIM and MRS fields of view in the JWST focal plane.  
The  axis is parallel to the along-slice axis of the MRS IFUs.}
\label{fig:mount}
\end{figure}

\clearpage

\begin{figure}[htbp]
\centerline{\includegraphics[width=5.0in]{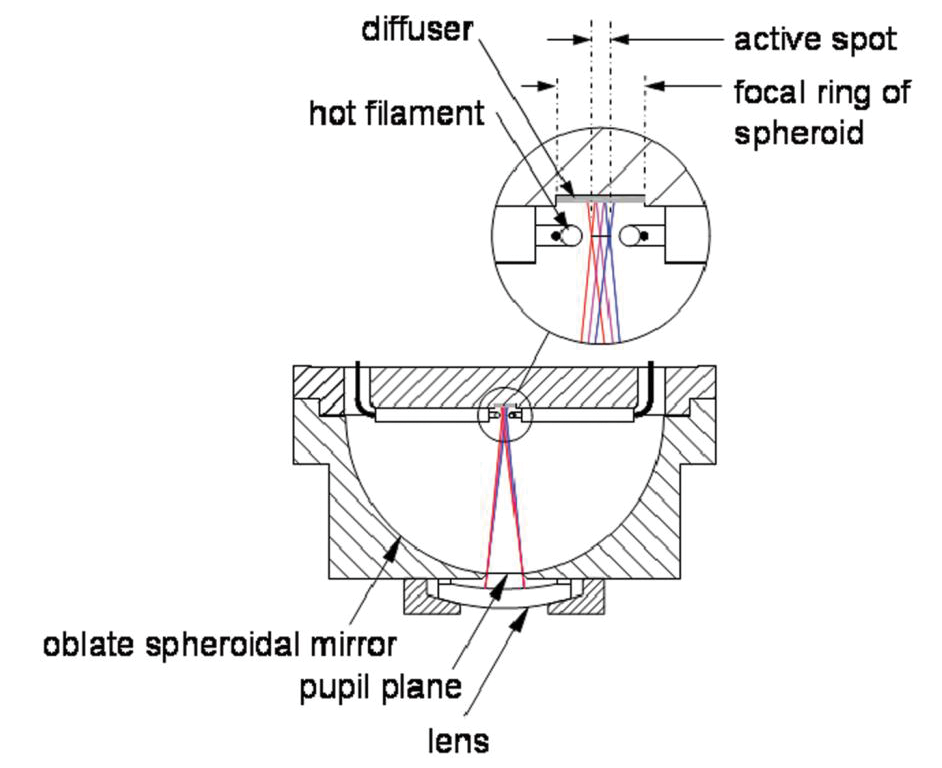}}
\caption{Calibration source. A hot tungsten filament illuminates a diffusing surface within an integrating sphere. Light escapes downward through the exit port of the integrating sphere. }
\label{fig:mount}
\end{figure}

\clearpage

\begin{figure}[htbp]
\centerline{\includegraphics[width=5.0in]{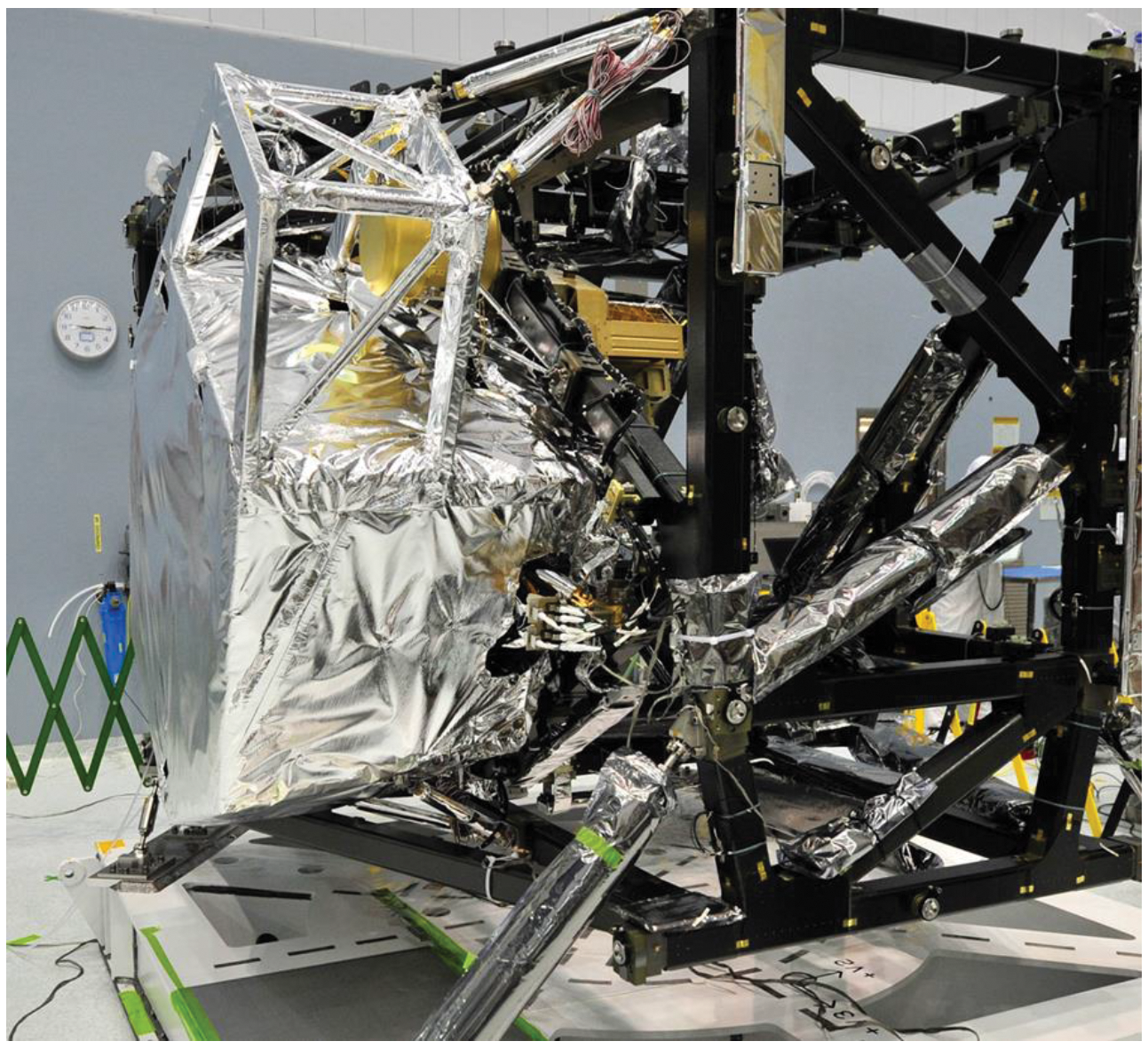}}
\caption{MIRI on ISIM and enveloped by the MIRI Thermal Shield, which provides a 23 K radiative environment.}
\label{fig:mount}
\end{figure}

\clearpage

\begin{figure}[htbp]
\centerline{\includegraphics[width=7.0in]{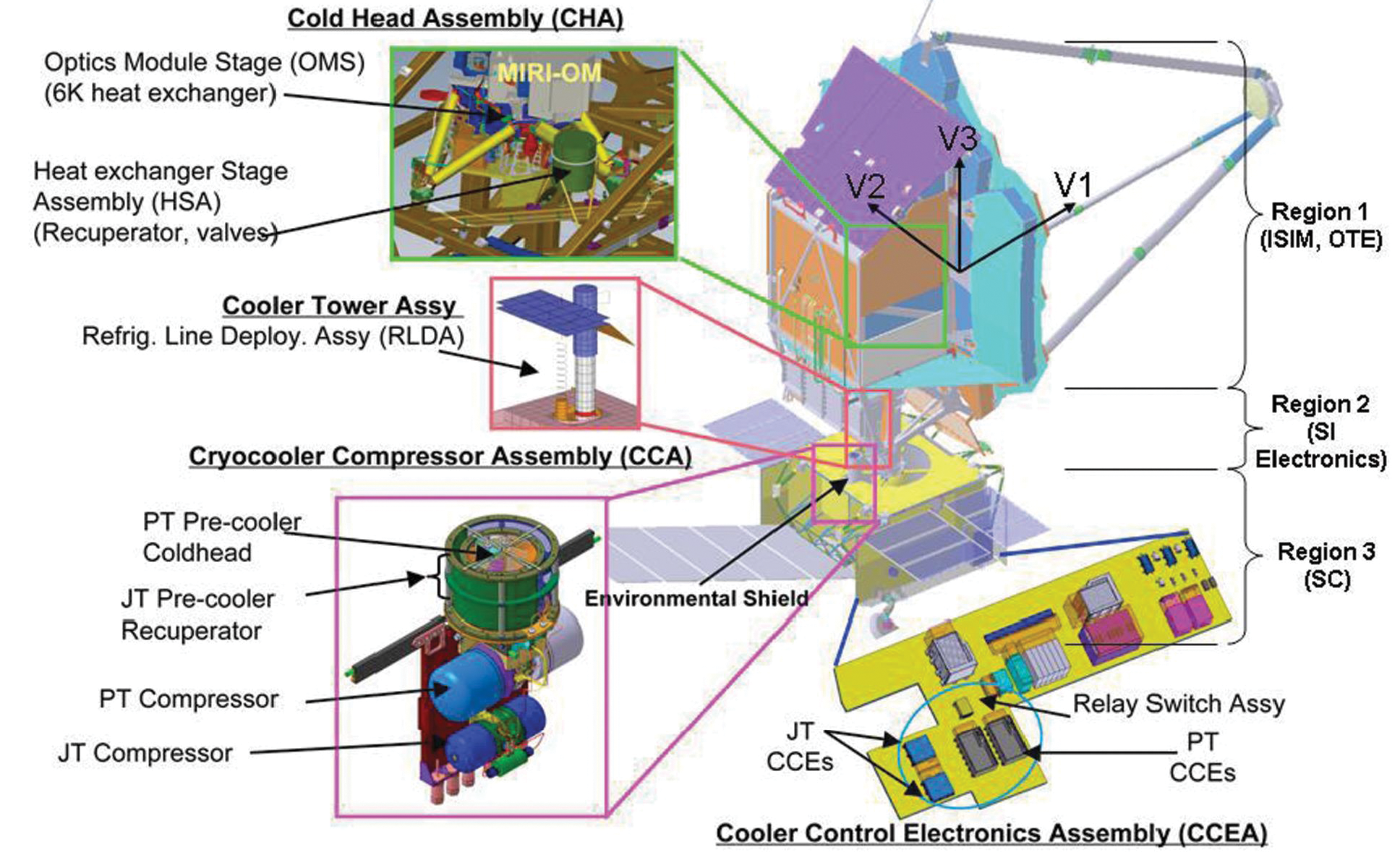}}
\caption{MIRI Cooler Components. From Banks et al. (2008).}
\label{fig:mount}
\end{figure}

\clearpage

\begin{figure}[htbp]
\centerline{\includegraphics[width=5.0in]{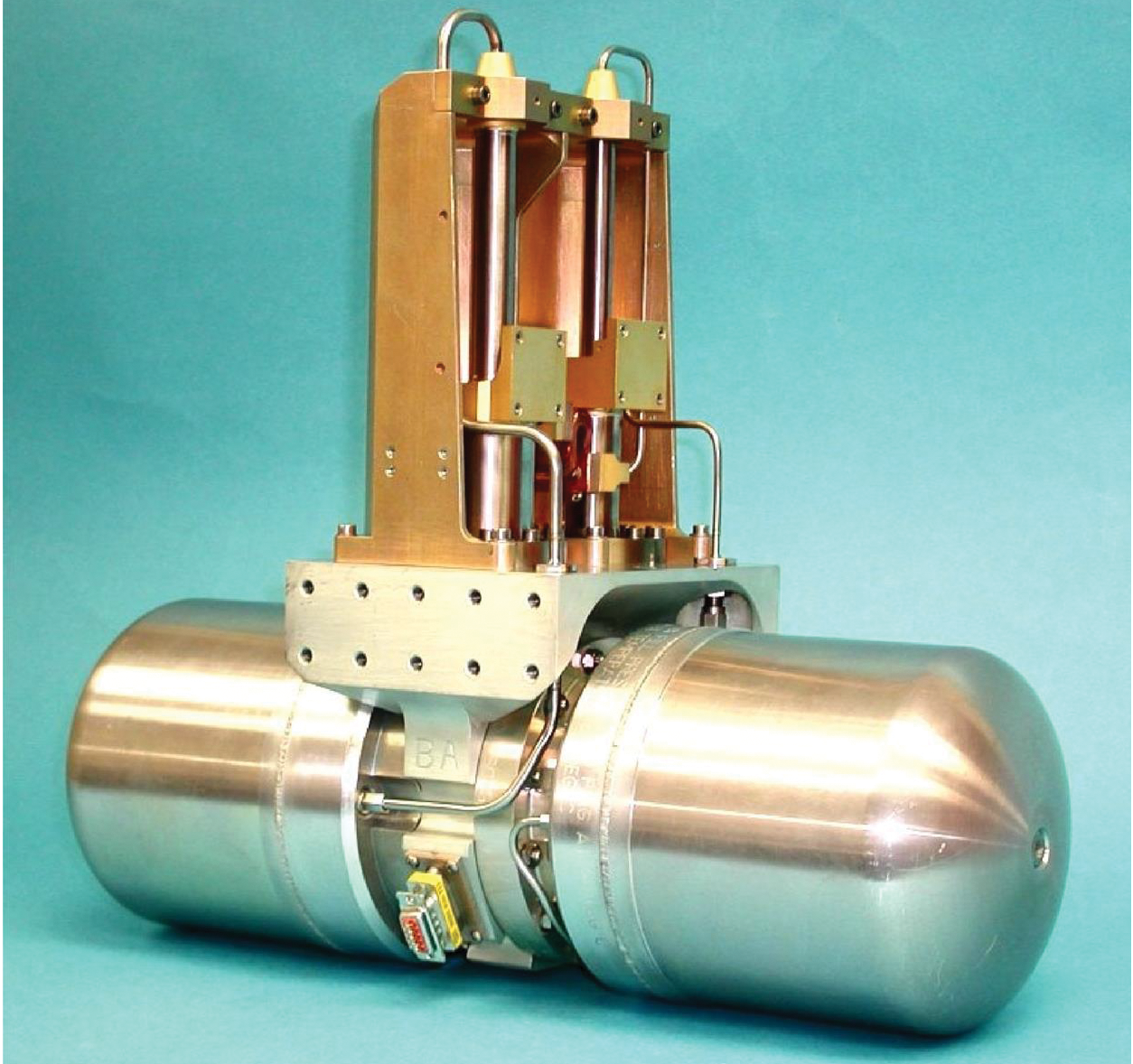}}
\caption{Pulse tube cooler (a predecessor of the MIRI flight model). The horizontal cylinders each contain a compressor; the two are driven in opposition to cancel vibration. The cold-stages of the three pulse tubes are connected thermally to the refrigerant line; the third stage is at ~ 18K. The helium gas is cooled to this temperature as it passes through to the RLDA, from which it is delivered to the optics module and to the Thermal OM shield.}
\label{fig:mount}
\end{figure}

\clearpage

\begin{figure}[htbp]
\centerline{\includegraphics[width=7.0in]{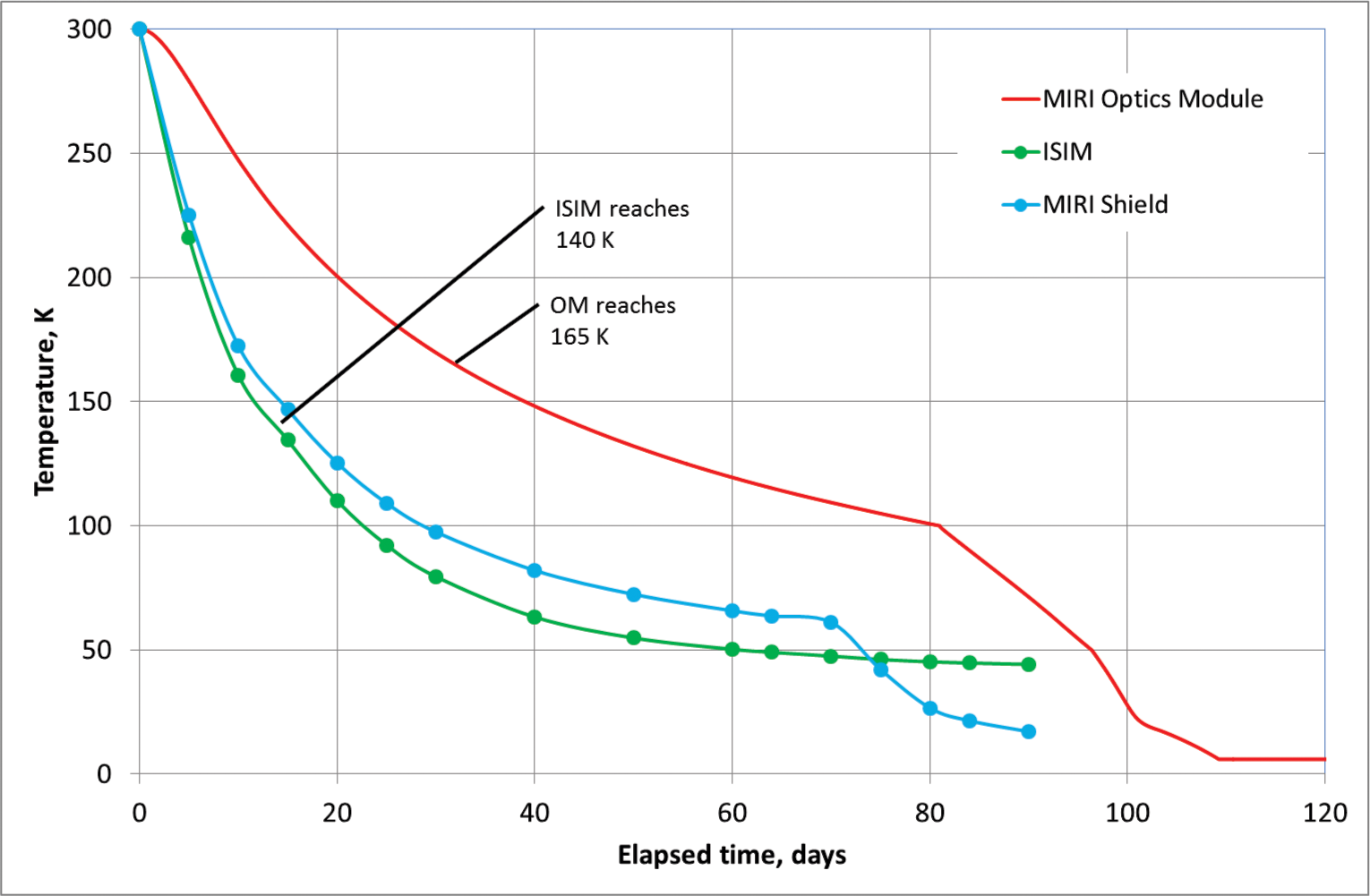}}
\caption{Cool-down prediction for MIRI}
\label{fig:mount}
\end{figure}

\clearpage

\begin{figure}[htbp]
\centerline{\includegraphics[width=5.0in]{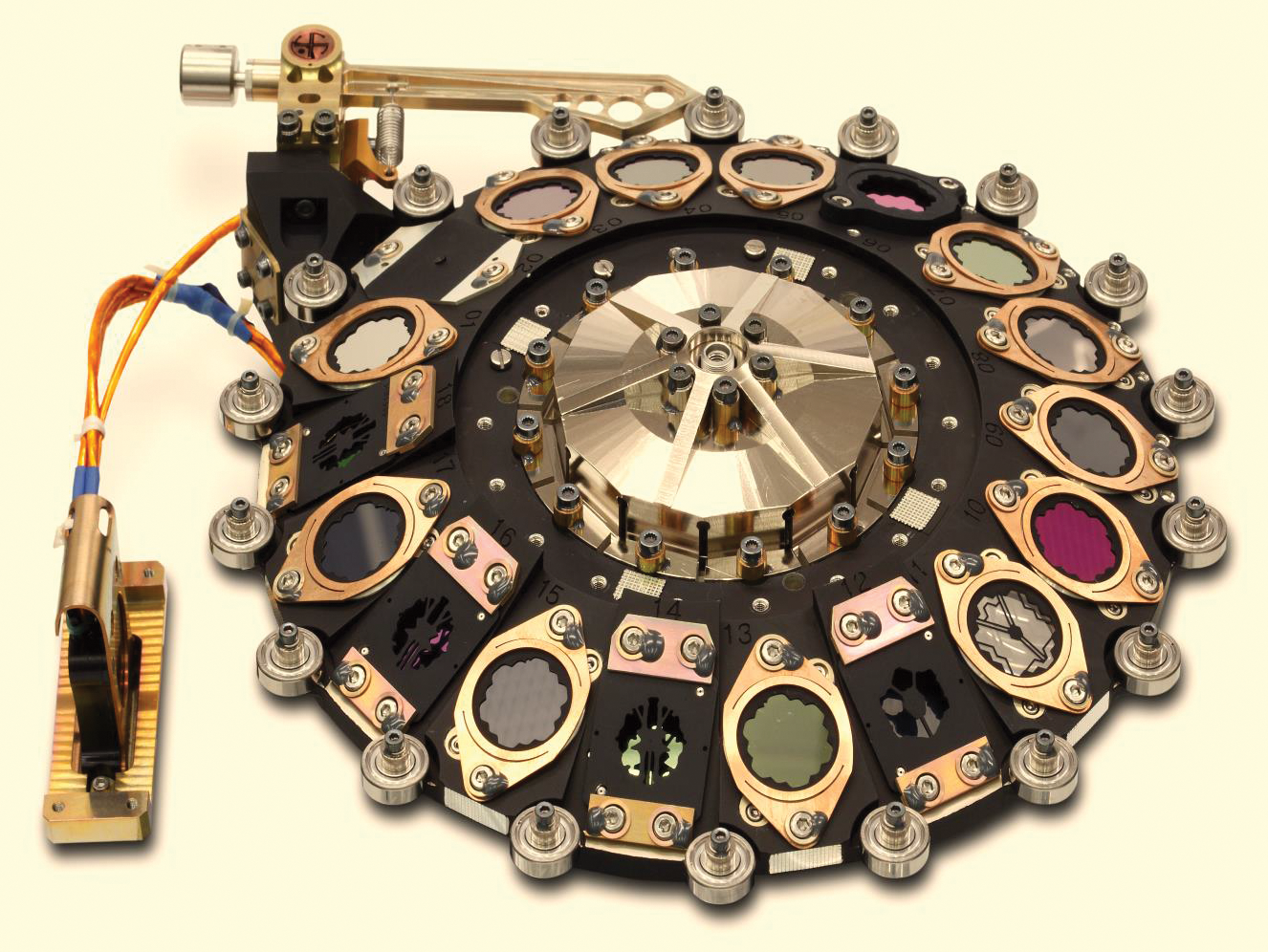}}
\caption{The MIRI Filter Wheel Assembly (FWA).}
\label{fig:mount}
\end{figure}

\clearpage

\begin{figure}[htbp]
\centerline{\includegraphics[width=5.0in]{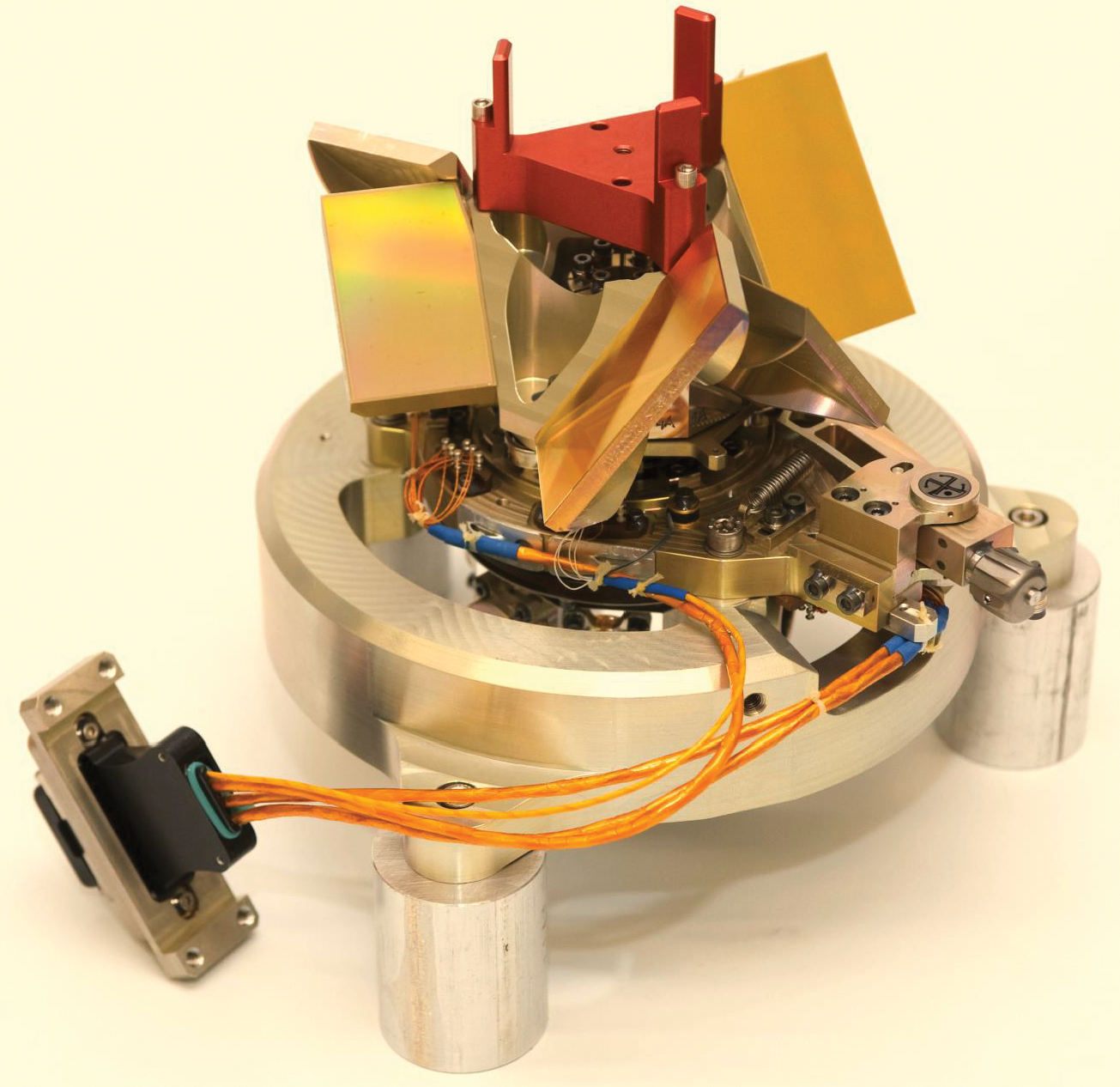}}
\caption{Dichroic/Grating Assembly A (DGA-A).}
\label{fig:mount}
\end{figure}

\clearpage

\begin{figure}[htbp]
\centerline{\includegraphics[width=5.0in]{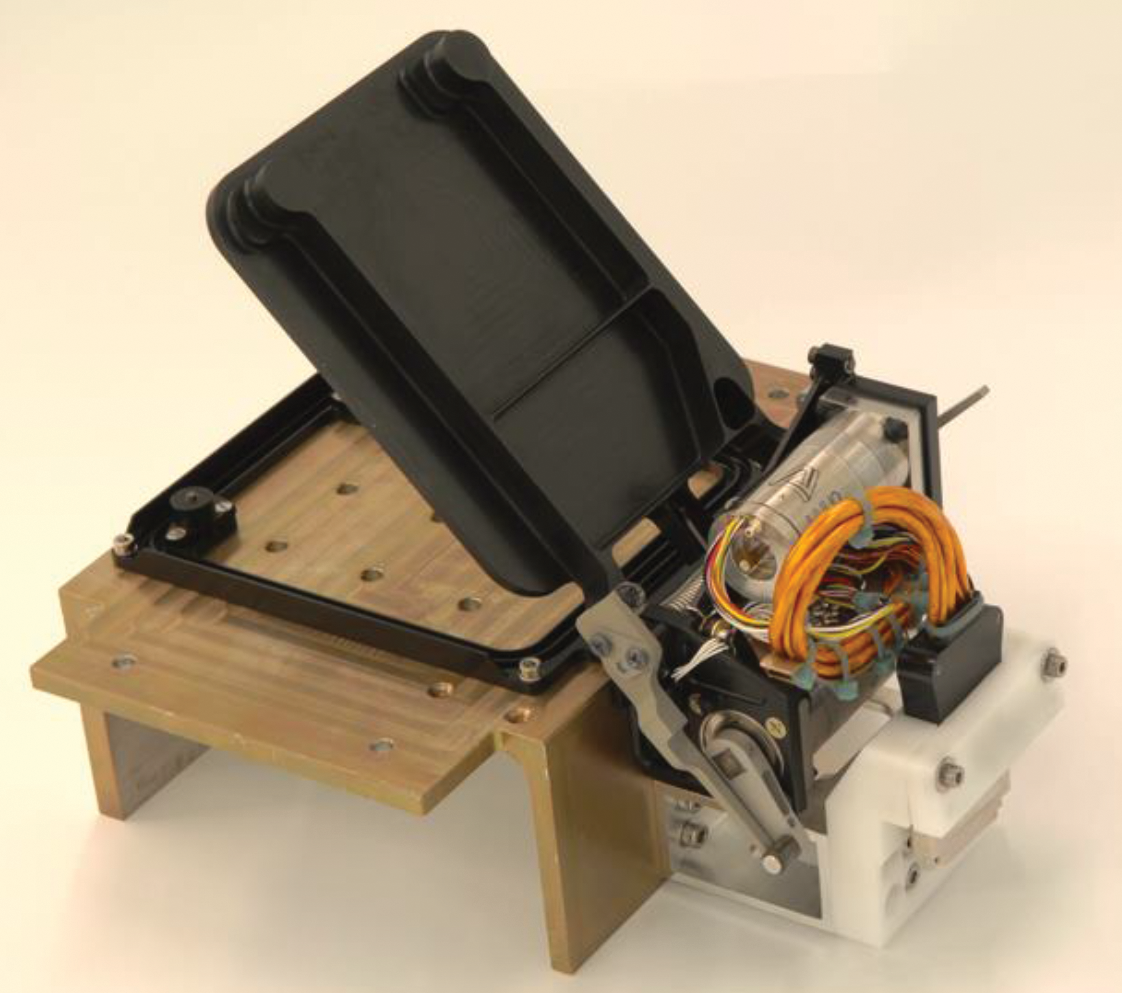}}
\caption{Contamination Control Cover (CCC) mounted on a mechanical support.}
\label{fig:mount}
\end{figure}

\clearpage

\begin{figure}[htbp]
\centerline{\includegraphics[width=5.0in]{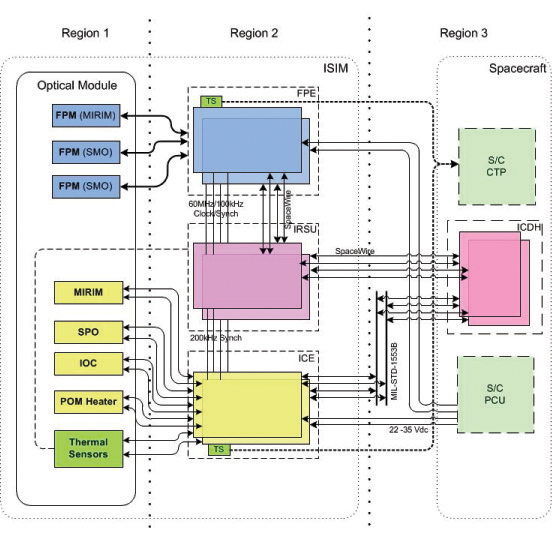}}
\caption{Electrical Architecture for the Optical System}
\label{fig:mount}
\end{figure}

\end{document}